\documentclass[longauth]{aa} 

\usepackage{graphicx}
\usepackage{txfonts}
\usepackage{svg}

\usepackage{hyperref}
\def\RJ{R_\mathrm{J}}
\def\MJ{M_\mathrm{J}}

\begin{document}

   \title{Water depletion and $\mathrm{^{15}NH_3}$ in the atmosphere of the coldest brown dwarf observed with JWST/MIRI} 

    \author{H. Kühnle\inst{1}  \and P. Patapis\inst{1} \and P. Mollière\inst{2} \and P. Tremblin\inst{3} \and E. Matthews\inst{2} \and  A. M. Glauser\inst{1} \and N. Whiteford\inst{4}  \and M. Vasist\inst{5} \and O. Absil\inst{5} \and D. Barrado\inst{6} \and  M. Min\inst{7} \and  P.-O. Lagage\inst{8} \and L. B. F. M. Waters\inst{7,9}  \and M. Guedel\inst{10,2,1} \and  Th. Henning\inst{2} \and B. Vandenbussche\inst{11}  \and P. Baudoz\inst{12} \and  L. Decin\inst{11}  \and  J. P. Pye\inst{13} \and P. Royer\inst{11} \and  E. F. van Dishoeck\inst{14} \and G. Östlin\inst{15} \and T. P. Ray\inst{16} \and G. Wright\inst{17} }
 
   \institute{
ETH Zürich, Institute for Particle Physics and Astrophysics, Wolfgang-Pauli-Str. 27, 8093 Zürich, Switzerland
\and
Max-Planck-Institut für Astronomie, Königstuhl 17, D-69117 Heidelberg, Germany
\and
Université Paris-Saclay, UVSQ, CNRS, CEA, Maison de la Simulation, 91191, Gif-sur-Yvette, France.
\and
Department of Astrophysics, American Museum of Natural History, New York, NY 10024, USA
\and
STAR Institute, Université de Liège, Allée du Six Août 19c, 4000 Liège, Belgium
\and
Centro de Astrobiología (CAB), CSIC-INTA, ESAC Campus, Camino Bajo del Castillo s/n, 28692 Villanueva de la Cañada, Madrid, Spain
\and
SRON Netherlands Institute for Space Research, Niels Bohrweg 4, 2333 CA Leiden, The Netherlands
\and
Université Paris-Saclay, Université Paris Cité, CEA, CNRS, AIM, F-91191 Gif-sur-Yvette, France
\and
Department of Astrophysics/IMAPP, Radboud University, PO Box 9010, 6500 GL Nijmegen, The Netherlands
\and
Department of Astrophysics, University of Vienna, Türkenschanzstr. 17, 1180 Vienna, Austria 
\and
Institute of Astronomy, KU Leuven, Celestijnenlaan 200D, 3001 Heverlee, Belgium
\and
LESIA, Observatoire de Paris, Université PSL, CNRS, Sorbonne Université, Université de Paris Cité, 5 place Jules Janssen, 92195 Meudon, France
\and
School of Physics \& Astronomy, Space Park Leicester, University of Leicester, 92 Corporation Road, Leicester, LE4 5SP, UK
\and
Leiden Observatory, Leiden University, P.O. Box 9513, 2300 RA Leiden, The Netherlands
\and
Department of Astronomy, Oskar Klein Centre, Stockholm University, 106 91 Stockholm, Sweden
\and
School of Cosmic Physics, Dublin Institute for Advanced Studies, 31 Fitzwilliam Place, Dublin, D02 XF86, Ireland
\and
UK Astronomy Technology Centre, Royal Observatory Edinburgh, Blackford Hill, Edinburgh EH9 3HJ, UK
}

   \date{Draft version October 14, 2024}

 
  \abstract
   {With a temperature of $\sim$285 K WISE~0855 is the coldest brown dwarf observed so far. Such cold gas giants enable probing atmospheric physics and chemistry of evolved objects similar to the Solar System gas giants. }
   {Using the James Webb Space Telescope (JWST) we obtained observations that allow us to characterize WISE~0855's atmosphere focusing on vertical variation in the water steam abundance, measuring trace gas abundances and receiving bulk parameters for this cold object.}
   {We observed the ultra cool dwarf WISE~0855 using the Mid-Infrared Instrument Medium Resolution Spectrometer (MIRI/MRS) onboard JWST at a spectral resolution of up to 3'750. We combined the observation with the published data from the Near Infrared Spectrograph (NIRSpec) G395M and PRISM modes yielding a spectrum ranging from 0.8 to 22 µm. We applied atmospheric retrievals using \texttt{petitRADTRANS} to measure atmospheric abundances, the pressure-temperature structure, radius and gravity of the brown dwarf. We also employed publicly available clear and cloudy self-consistent grid models to estimate bulk properties of the atmosphere such as the effective temperature, radius, gravity and metallicity.}
   {Atmospheric retrievals constrain a variable water abundance profile in the atmosphere, as predicted by equilibrium chemistry. We detect the $\mathrm{^{15}NH_3}$ isotopologue and infer a ratio of mass fraction of $\mathrm{^{14}NH_3}$/$\mathrm{^{15}NH_3}$ = 332$^{+63}_{-43}$ for the clear retrieval. We measure the bolometric luminosity by integrating the presented spectrum and obtain a value of $\mathrm{log(L/\textup{L}_\odot)} = -7.291 \pm 0.008$.}
   {The detected water depletion indicates that water condenses out in the upper atmosphere due to the very low effective temperature of WISE~0855. The height in the atmosphere where this occurs is covered by the MIRI/MRS data, and thus demonstrates the potential of MIRI to characterize cold gas giant's atmospheres. Comparing the data to retrievals and self-consistent grid models, we do not detect signs for water ice clouds, although their spectral features have been predicted in previous studies. }

   \maketitle
%

\section{Introduction}

Y dwarfs are characterized by cold temperatures below 500 K \citep{Cushing2011}. As they are often either far from their host star or free-floating, they are prime targets for spectroscopic characterization with direct imaging. Their atmospheres are comparable to gas giant's atmospheres in terms of their temperature and composition \citep{Burrows1997,Beichman2014}. In addition, they can be used to characterize evolved atmospheres and investigate the formation of weather patters and clouds similar to the processes in Jupiter in the Solar System \citep{Coulter2022}. For the first time, the \textit{James Webb Space Telescope (JWST)} \citep{Gardner2023,Rigby2023} enables to measure and characterize such cold objects in the mid-infrared. \citet{Barrado2023} presented the first detection of the ammonia isotopologue $^{15}$NH$_3$ in the Y dwarf WISEPA J182831.08+265037.8 (hereafter WISE~1828). In the near-infrared, \citet{Faherty2024} find methane emission in the isolated Y dwarf CWISEP J193518.59-154620.3 which could be linked to a temperature inversion in the upper atmosphere possibly originating from aurorae. A sample of Y dwarf atmospheres has been studied using JWST presenting their bolometric luminosities and effective temperatures \citep{Beiler2024a}. In regards to the temperature of WISE~0855, the only comparable planetary-mass object found to date is the recently detected $\epsilon$~Indi~Ab by \citet{Matthews2024} with a temperature of about $\sim$ 275 K. This exciting detection might be the closest Jupiter analog observed so far. 

\citet{Luhman2014b} detected WISEJ0855–0714 (hereafter WISE~0855) and classified it as a late Y dwarf in WISE and Spitzer measurements. It was identified as the coldest brown dwarf from its low flux at 4.5 µm and reddest in [3.6] - [4.5] color. It is the fourth closest object to our Solar System with a parallax of 438.9 $\pm$ 3.0 mas and corresponding to a distance of 2.28 $\pm$ 0.02 pc \citep{Kirkpatrick2019}. The age has not been precisely determined with a rough estimate of 0.3-6 Gyr and a mass of 1.5-8 M$_J$ when comparing to evolutionary models \citep{Leggett2017}. WISE~0855 is an isolated brown dwarf and thus a perfect laboratory for direct imaging and atmospheric characterization through spectroscopy without the need of Point Spread Function (PSF) subtraction procedures from stellar light contamination. 
\citet{Skemer2016} presented the first spectrum of WISE~0855 in the M band (4.5–5.2 $\mu$m) using the Gemini-North telescope and the Gemini Near Infrared Spectrograph (GNIRS). \citet{Luhman2016} added more photometric data from HST, Spitzer and Gemini North. Follow up observations studying the variability of WISE~0855 were performed using the Spitzer telescope in the Spitzer/IRAC I1 and I2 filters centered at 3.6 and 4.5 µm \citep{Esplin2016}. \citet{Morley2018} added an L band (3.4 to 4.14 $\mu$m) spectrum as well from GNIRS. The new era of JWST and specifically the \textit{Mid-Infrared Instrument (MIRI)} \citet{Wright2015,Wright2023}) allows us to study cold Y dwarfs as they peak in flux far in the mid-infrared. Covering the wavelengths from 5 to 28 µm in resolving power of up to R $\approx$ 3'750 the \textit{Medium Resolution Spectrometer (MRS)} \citep{Wells2015,Argyriou2023} allows us to characterize their atmospheric physics and chemistry. The recent publication of \citet{Luhman2024} presents JWST/NIRSpec PRISM and G395M/F290LP spectroscopic data of WISE~0855 encompassing a wavelength region of 0.8 to 5 $\mu$m.

Due to the low temperatures of WISE~0855, water is expected to condense out in the atmosphere. Theoretical modelling approaches argue either for or against the existence of water ice clouds in brown dwarfs with temperatures below $T_{\rm eff}\leq$ 350 K \citep{Burrows2003, Morley2012,Esplin2016,Mang2022}. \citet{Morley2014} presents the scattering albedo of water ice clouds showing strong features at 2.8 $\mu$m and 10 $\mu$m - both encompassed in the NIRspec and MIRI wavelength regions respectively. 
\citet{Faherty2014} reported the first indication of water ice clouds in WISE~0855 by comparing the WISE photometry with the models developed by \citet{Morley2014} and \citet{Saumon2012}. \citet{Skemer2016} compared the M band spectrum to models finding better agreement with cloudy models and \citet{Morley2018} shows evidence for water ice clouds when comparing both the L and M band spectra to cloudy and clear models. \citet{Lacy2023} compare their models to earlier observations and model simulations favouring thick water clouds and non equilibrium in WISE~0855. 
However, \citet{Luhman2014b} showed that the same measurements can be explained with cloudless models including non-equilibrium chemistry. \citet{Esplin2016} presented several reasons leading to the observed variability in the Spitzer data, including but not only related to the presence of water clouds. \citet{Tremblin2015} show that another process could explain especially the lower flux in the NIR by changing the slope of the Pressure-Temperature (PT) structure in the lower atmosphere. They identified fingering convection to be a physical process leading to a cooling of the lower atmosphere and thus to such an adaption of the PT structure. Further, rain-out condensation similar to the effect observed in L- and T-dwarfs presented in \citet{Marley2002} might be a process in which condensates sink quickly below the photosphere and do not form stable clouds due to the higher surface gravity of brown dwarfs compared to planets making them unobservable. \citet{Luhman2024} fit the JWST/NIRspec data well when comparing it to the clear ATMO++ models which implement this change in the PT structure. 
\citet{Tremblin2019} generalize their theory about fingering convection showing that the same physical principle is present in the thermohaline convection in Earth's oceans as well as in chemical processes resulting in CH$_4$ and CO abundances out of equilibrium.

Disequilibrium chemistry is expected in cold brown dwarf atmospheres \citep{Saumon2000,Zahnle2014,Miles2020,Beiler2024b}. Non-equilibrium effects denote an atmosphere influenced by processes such as mixing, convection, or zonal jets, leading to a constant replenishment of the chemical compounds before chemical reactions can modify them. The chemical timescale is thus larger than the dynamical timescale \citep{Zahnle2014}.
CO has previously been found in WISE~0855 \citep{Miles2020}, hinting at dynamical effects in the atmosphere. CO is found in deep and hot layers in the atmosphere of cold brown dwarfs, however it can be transported to upper layers by vertical mixing. Thus, compared to chemical equilibrium, we expect to find in disequilibrium a higher CO abundance together with less absorption by $\mathrm{CH_4}$ in the 4-5 $\mu$m range \citep[e.g.][]{Saumon2006,Morley2014,Miles2020}.  
A similar effect is predicted by a decrease in $\mathrm{NH_3}$ and an increase in $\mathrm{N_2}$ in the upper atmosphere of cold brown dwarfs. While $\mathrm{NH_3}$ has characteristic absorption lines in the 10-14 $\mu$m range, $\mathrm{N_2}$ is difficult to observe \citep{Lodders2002,Saumon2006}. Together CO, CH$_4$ and NH$_3$, along with PH$_3$, allow for the inference of chemical disequilibrium. $\mathrm{PH_3}$ has not been detected so far in WISE~0855 even though it is highly abundant in Jupiter. The reason for missing $\mathrm{PH_3}$ is not yet fully understood \citep{Skemer2016,Morley2018,Luhman2024}. The recent publication by \citet{Beiler2024b} presents pathways how the PH$_3$ may be depleted in the atmospheres of late T and Y dwarfs. Incomplete theoretical understanding, namely missing chemical reactions in the phosphorous chemistry, and potential condensation of $\rm NH_4H_2PO_4$ in the lower atmosphere might lead to a lower PH$_3$ abundance.

Nitrogen isotopologues in Y dwarf atmospheres have been shown in MIRI/MRS data of WISE~1828, thanks to the spectrally resolved lines in the mid infrared \citep{Barrado2023}. Isotopologue ratios are introduced as a new formation tracer to study the formation of gas giants and brown dwarfs while being studied already on Solar System bodies. Isotopologues help to understand the role of the nitrogen isotopes in the formation history of planetary bodies \citep{Adande2012,Zhang2021,Nomura2022}. To better understand the fractionation and formation scenarios, more isotopologue ratios need to be measured. 

Atmospheric free retrievals have become an important tool to characterize atmospheres of gas giants \citep[e.g.,][]{Madhusudhan2009, Benneke2012,Molliere2020}. This allows measuring the composition, abundances and the atmospheric thermal structure. Using a forward model encompassing a radiative transfer code many spectra are simulated for input parameters sampled in defined prior distributions.

Self-consistent radiative-convective grid models help characterizing gas giant atmospheres building on first principles and taking processes and interactions of atmospheric chemistry and physics into account \citep{Phillips2020,Leggett2021,Lacy2023,Mukherjee2024}. 
A recent model grid by \citet{Lacy2023} introduces water ice clouds to cold brown dwarf atmospheres. Using previous observations and model simulations, \citet{Lacy2023} conclude that in the W3 filter region ($\sim$ 12 $\mu$m) and in the M (4.5–5.2 $\mu$m) band vertically extended cloud models fit best. 
As presented in \citet{Leggett2021} and \citet{Meisner2023} the ATMO++ models are compared to simulated Y dwarf spectra and observations. This set of model grids is an improved version of the cloudless and disequilibrium model ATMO++ without PH$_3$ \citep{Phillips2020} for lower temperature brown dwarfs by adjusting the temperature pressure gradient. \citet{Mukherjee2024} present the Sonora Elf Owl models which explores the disequilibrium chemistry and dynamics in giant exoplanets and brown dwarfs.

In this work, we present the MIRI/MRS spectrum of the coldest yet observed brown dwarf and characterize its atmosphere using atmospheric retrievals and grid models. We present the obtained data set and its data reduction in Section \ref{data}. The analysis methods, namely the atmospheric retrievals as well as the self-consistent grid models are presented in Section \ref{method}. In Section \ref{results}, we present our results of the analysis and in Section \ref{discussion} we discuss them. Section \ref{conclusion} concludes the findings of this work and presents an outlook. 

\begin{figure*}[h] 
   \centering
   \includegraphics[width=1\textwidth]{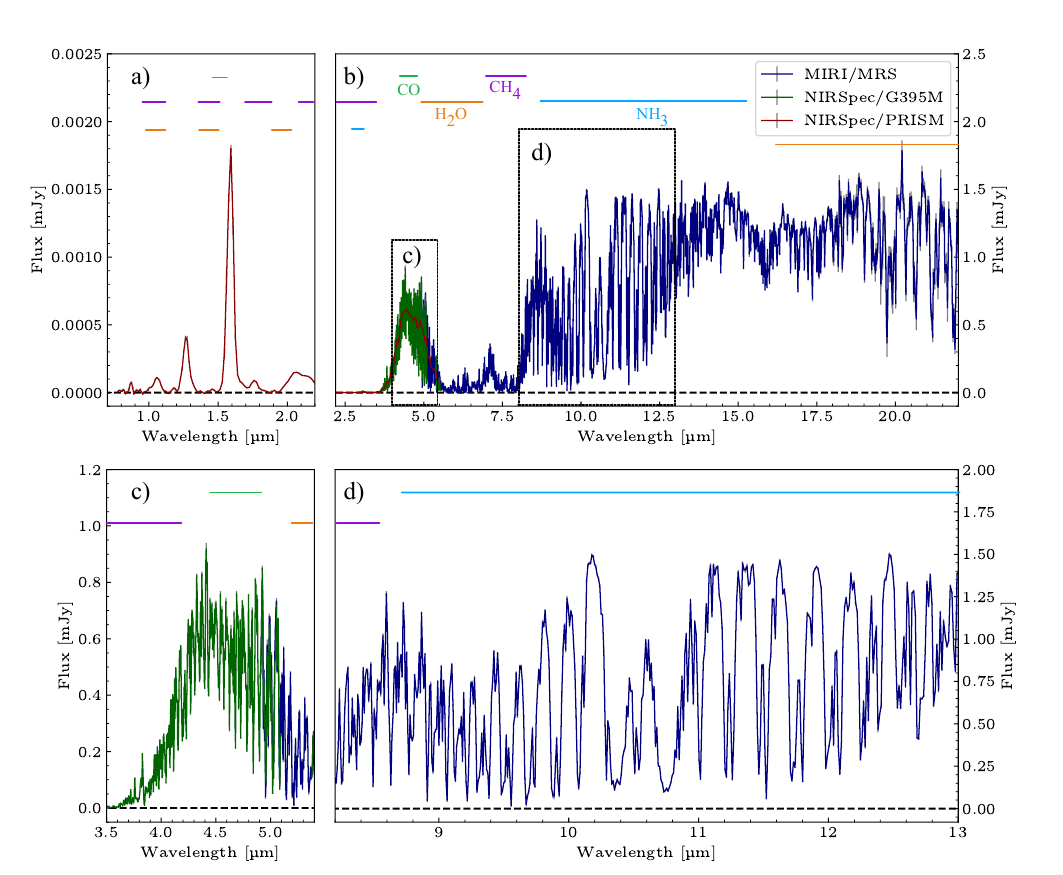}
   \caption{The MIRI/MRS (this work), NIRspec/PRISM and NIRSpec/G395M \citet{Luhman2024} spectrum of WISE~0855 from 0.8 to 22 µm.  Panel a) shows the NIRspec/PRISM data in red and panel from 0.8 to 2.2µm b) the NIRspec/G395M in green and MIRI/MRS data in blue from 2.2 to 22 µm. Insets c) and d) show the overlap between the three data sets from 3.5 to 5.4 µm and the ammonia feature from 8.0 to 13.0 µm, respectively. The most dominant absorbing species are presented above the spectrum in various colors. For better visualization he data was binned to a resolution of R = 1000 for MIRI/MRS and NIRSpec/G395M and R = 100 for NIRSpec/PRISM.}%
   \label{fig:spectrumfull}
\end{figure*}

\section{Observation and data reduction} \label{data}

    \begin{figure*}[h] 
   \centering
   \includegraphics[width=1\textwidth]{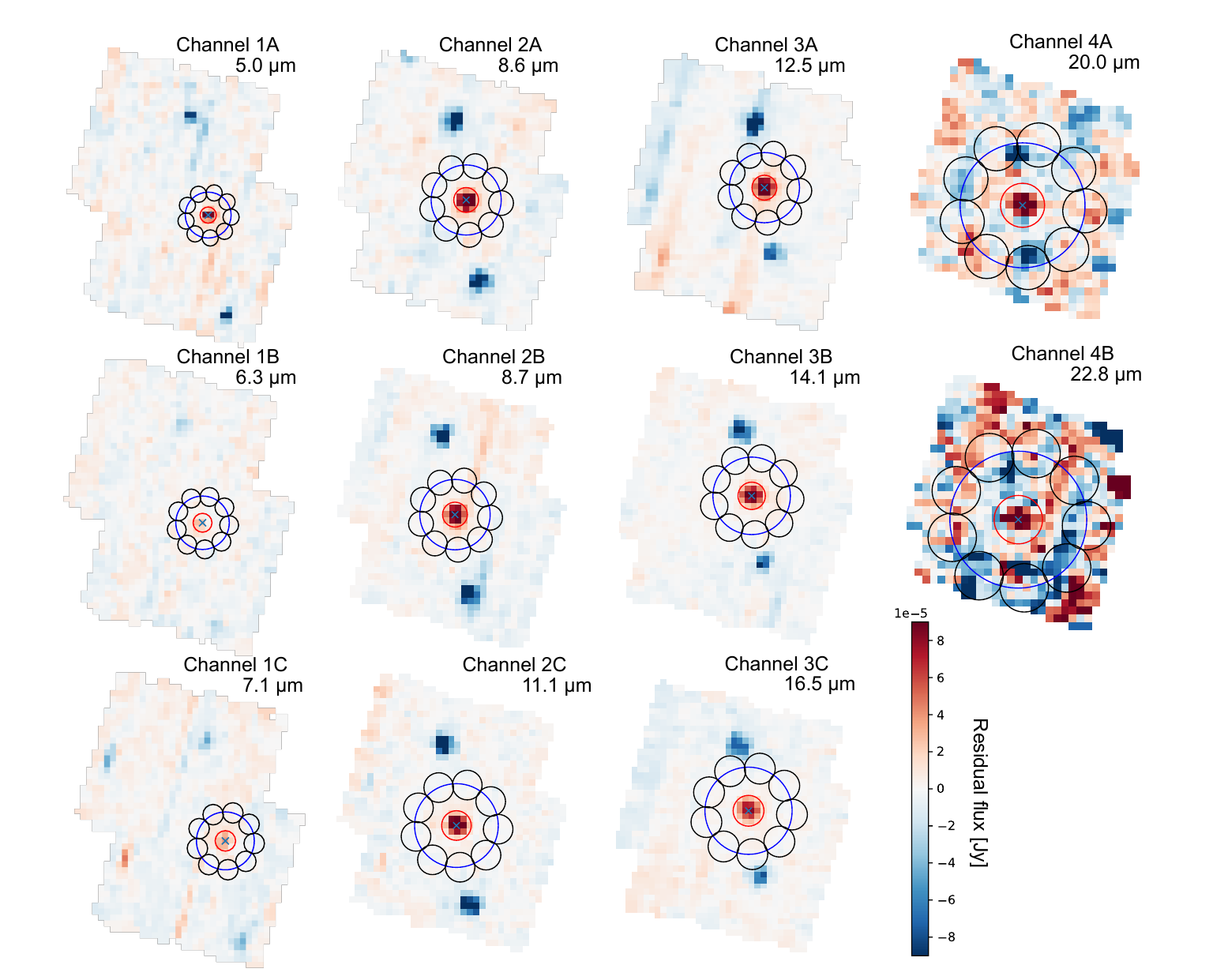}
   \caption{Resulting cube images of channels 1 to 4 with sub channels A to C after dither subtraction. The wavelengths are chosen such that they refer to the detector image with the highest flux at the source in each band. The red circle shows the aperture for the flux extraction of the spectrum at one FWHM of the PSF of the source. The blue circle refers to three times the FWHM on which nine black circles of one FWHM are placed. We extract the flux from each circle and measure the variability to obtain an error including the noise in the background, $\sigma_{emp}$. }%
    \label{fig:errorestimation}
    \end{figure*}
    
    WISE~0855 was observed on the 17th of April, 2023 with MIRI/MRS onboard JWST. The observation is part of the MIRI European Consortium Guaranteed Time Observations (GTO) program 1230 (PI: C. Alves de Oliveira).
    The effective exposure time was 993.5~s and the target was observed at the position RA = 133.77 $\pm$ 0.04 deg and DEC = -7.24 $\pm$ 0.04 deg during the mid time of the observation.
    The observation was run in fast readout pattern FASTR1 in a 2-point dither pattern, and one integration of 179 groups.

    The data were reduced using the JWST pipeline version 1.12.5 for the reduction of stages one to three \footnote{https://jwst-pipeline.readthedocs.io/en/latest/} \citep{Bushouse2023}. The CRDS files used for the presented spectrum are of version jwst\_1149.pmap. 
    Stage one generates rate files after correcting for the dark current and other detector effects. Stage two applies the pipeline built-in flat field, stray light, fringe and photometric correction on the rate files.
    We reduced noise in the background by subtracting the two dithers from each other after the second stage similar to what was done in \citet{Barrado2023}. As the two dithers were acquired shortly after each other, the background stays similar across both acquisitions. To build the cube the pipeline uses the \textit{drizzle} weighting algorithm \citep{Law2023} in the third stage. 
    From the cube we extract the 1d spectrum by applying an aperture of a radius of 1.0 $\times$ FWHM of the PSF for the respective wavelength on the source pixel. Due to the faint source we avoid introducing additional noise by choosing a small aperture. The coordinates of the source are selected by the built-in \texttt{ifu\_autocenter()} function. For channel 1B the algorithm could not detect the source by itself and we centered the aperture manually to the pixel with the highest intensity in the summed cube over the respective channel. The built-in residual fringe correction \texttt{ifu\_rcorr()} in the 1d spectrum is applied. 
    In Figure \ref{fig:spectrumfull} we present the resulting spectrum of WISE~0855 from 0.8 to 22 µm by combining the MIRI/MRS with the NIRspec/PRISM and NIRspec/G395M grating data presented in \citet{Luhman2024}.%
    
    MIRI/MRS shows a channel dependent resolution from R $\sim$ 1'500 in channel 4 to R $\sim$ 3'750 in channel 1 \citep{Jones2023}. As MIRI/MRS consists of an Integral Field Unit (IFU) we can display the data in two spatial and one wavelength dimension, in a so-called image cube. Figure \ref{fig:errorestimation} shows the image cubes of each subchannel at the wavelength with highest flux at the source per channel. The red circle corresponds to the aperture set around the source with a radius of 1.0 $\times$ FWHM of the PSF and the cross indicates the source center pixel. The FWHM of the PSF for MIRI/MRS is presented in \citet{Law2023}. In Figure \ref{fig:errorestimation} we see that channel 4B already is strongly dominated by noise and it becomes even worse for channel 4C (not depicted). Therefore, we neglect channel 4C and parts of channel 4B for the analysis presented in this work.  
    
    The NIRSpec/G395M grating has a resolution of R $\sim$ 1000 and NIRSpec/PRISM of R $\sim$ 30-300 \citep{Boker2023} and the data were acquired in the context of the same GTO program (1230) on the same date. Further acquisition and data reduction specifications of the NIRspec data are presented in \cite{Luhman2024}.

    The error estimated by the pipeline is typically small in the order of 0.5\% of the spectrum as it is presented in Figure \ref{fig:errors}. In addition, as shown in Figure \ref{fig:errorestimation} the residual background shows a slight striping effect depending on the wavelength especially pronounced for 8.7 and 12.5 µm in this Figure. As this should be reflected accordingly in the error on the data we use the following methodology.
    For every wavelength we place a series of 9 noise apertures at a distance of 2.0 $\times$ FWHM from the source center, each with a radius of 1.0 $\times$ FWHM. The used apertures on selected wavelengths around the source are shown in Figure \ref{fig:errorestimation}. In the next step we take the 68th percentile of this distribution as the wavelength dependent additional error by adding it in quadrature to the pipeline error, this error being referred to as empirical error. As we only take nine apertures, we multiply the error by a factor penalizing small sample sizes as shown in \citet{Mawet2014}. This procedure results in larger errors for larger variability in the background. Also by taking the percentile compared to the standard deviation, we are more conservative. With this method we however assume random noise following a Gaussian distribution.  This procedure was only used for the MIRI/MRS data. The error estimate is compared to the pipeline error in Figure \ref{fig:errors} in the appendix.

    In addition to the empirical error, we increase the error on all data sets by a certain offset. We use the approach described in \citet{Line2015} by defining a 10$^b$ factor which is added to the squared uncertainty to account for any unknown uncertainty. The parameter $b$ is then retrieved as a free parameter in the atmospheric retrieval from a prior set between the minimum and maximum error on the data. In this analysis we use a different value for the NIRspec/PRISM, NIRspec/G395M and MIRI/MRS data sets. For NIRSpec/PRSIM we use two values for wavelengths smaller and larger than 2.2µm due to the large absolute flux difference between the two wavelength ranges. For MIRI/MRS the squared uncertainty already includes the previously described empirical error correction. The retrieved values of the clear retrieval are then used for the error inflation in the self-consistent grid model analysis. 

    The final error for the MIRI/MRS data $\sigma_{\rm MIRI}$ used in the analysis consists of three different error components, the pipeline error $\sigma_{\rm pipe}$, the empirical error $\sigma_{\rm emp}$ and the retrieved error inflation factor 10$^b$ as explained above:
    \begin{equation}
        \sigma_{\rm MIRI}^2 = \sigma_{\rm pipe}^2 + \sigma_{\rm emp}^2 + 10^b
    \end{equation}
    For NIRSpec/G395M and NIRSpec/PRISM we have not calculated an empirical error, but have added the retrieved inflation factor as well. Figure \ref{fig:errors} in the appendix compares the two error corrections with the pipeline error. Figure \ref{fig:snr} presents the corresponding Signal to Noise Ratio (S/N) for both error estimates. The SNR for the empirical error with the inflation factor is in the order of 5 to 10, which seems plausible. However, the usage of the rather largely inflated error may have implications, such as yielding larger uncertainties on the fit results and larger posterior distributions with the nested sampling algorithm. 

\section{Methodology} \label{method}

    \subsection{Atmospheric Retrievals} \label{met:atmoret}
    To understand the atmospheric structure and obtain abundance estimates, we perform a free atmospheric retrieval analysis using the publicly available python package \texttt{petitRADTRANS} \cite{Molliere2019,Nasedkin2024} (version 2.7.4).  We use the sampling algorithm \texttt{PyMultinest} \citep{Buchner2014} which is based on the nested sampling algorithm \texttt{MultiNest} \citep{Skilling2004,Feroz2007}.
    
    We present the results of three different retrievals. The following parameters are in common for all retrievals: The logarithmic mass fractions of $\mathrm{H_2O}$, $\mathrm{NH_3}$, $\mathrm{^{15}NH_3}$, $\mathrm{CH_4}$, $\mathrm{CO}$, $\mathrm{H_2S}$ and $\mathrm{PH_3}$ gas (assumed to be vertically constant in the atmospheric column), the planet radius $\mathrm{R_p}$, gravity $\mathrm{log~g}$, the base temperature $\mathrm{T_{bottom}}$ as well as nodes in the temperature structure, where the profile can be adapted. The baseline retrieval assumes a clear atmosphere. To test whether the abundance of water in the gas phase may decrease with altitude even without adding water ice clouds, we perform a second retrieval for a clear atmosphere with a variable water profile with pressure, by adding two free parameters: A pressure level chosen freely throughout the column, $\mathrm{p_{H_2O}}$, and an exponent of a power law reducing the initial water mass mixing ratio $\alpha$. This allows for a decrease or increase at lower pressures compared to the initial water abundance parameterized as shown in Eq. \ref{eq:waterparam}:
    \begin{equation} \label{eq:waterparam}
        \rm M(\rm H_2O)(p) = M(\rm H_2O)_{\rm r}\left(\frac{p}{p_{H2O}}\right)^{\alpha}, \ {\rm for} \ p < p_b,
    \end{equation} where $\mathrm{M(H_2O_{cd})(p)}$ denotes the logarithmic mass fraction of water vapour varying with pressure and $\mathrm{M(H_2O_{cd})_{r}}$ the retrieved water vapour value for the lower atmosphere.

    Increasing the complexity in the model further, we add clouds in a third retrieval whilst keeping the variable water profile of the second retrieval. Here, we include the water cloud opacities $\mathrm{H_2O_{cd}}$ corresponding to spherical ice particles as well as the following cloud parameters: The sedimentation coefficient $f_{\rm sed}$, the logarithmic eddy mixing coefficient $\mathrm{log K_{zz}}$, the width of the particle size distribution $\mathrm{\sigma_{lnorm}}$ and the logarithmic pressure at the lower cloud base $\mathrm{logP_{base}}$. The amount of cloud particles decreases with height according to a freely retrieved parameter $f_{\rm sed}$. Small values for $f_{\rm sed}$ correspond to large extensions of the cloud deck and larger $f_{\rm sed}$ values to smaller cloud extensions. The cloud layer is given by the following parameterization based on the work of \citet{Ackerman2001} presented in Eq. \ref{eq:cloudparam}, where $\mathrm{M(H_2O_{cd})(p)}$ corresponds to the logarithmic mass fraction of condensate molecules in the cloud layer varying with pressure, $\mathrm{M(H_2O_{cd})_{r}}$ to the retrieved value of the condensate abundance and $\rm p_b$ to the retrieved cloud base layer pressure. In this context, 'cd' stands for crystalline and 'distribution of hollow spheres' (DHS) particles approximating a non-spherical shape of crystals \citep{Min2005}. The priors and posteriors of the three retrievals are presented in the appendix in Table \ref{table:posteriors}. 
    \begin{equation} \label{eq:cloudparam}
        \rm M(\rm H_2O_{cd})(p) = M(\rm H_2O_{cd})_{\rm r}\left(\frac{p}{p_b}\right)^{f_{\rm sed}}, \ {\rm for} \ p < p_b.
    \end{equation}
    
    To detect the isoptopologue $\mathrm{^{15}NH_3}$ we perform one additional retrieval, being identical to the clear retrieval with constant water profile, but neglecting the opacity of $\mathrm{^{15}NH_3}$ and the associated free parameter for the $^{15}{\rm NH}_3$ abundance. 
    
    We use correlated k (c-k) opacities at a wavelength binning of $\lambda / \Delta \lambda = 1000$. Thus, we bin the MIRI/MRS and NIRspec/G395M data to the corresponding wavelength grid of petitRADTRANS. The NIRspec/PRISM data set has a resolution of R $\sim$ 100 and is retrieved on this resolving power simultaneously with the higher resolution NIRspec/G395M and MIRI/MRS data sets. The used opacities for the retrievals are shown in the appendix in Figure \ref{fig:opacities}. 
    
    A forward model simulates one dimensional emission spectra from a PT profile parameterized depending on an interior temperature and a custom number of nodes, as well as a given set of opacities and their abundances. The PT structure is parameterized using ten nodes, a bottom temperature and a spline interpolation between the nodes. Starting from the bottom temperature the nodes define the factor by which the temperature is changed for the next layer. We choose the nodes in a prior range between 0.2 and 1.0 and the bottom temperature between 100 K and 9000 K. To better constrain the temperature profile we added a freely retrieved regularization term depending on a penalty parameter $\gamma$ \citep{Line2015}. For large values of $\gamma$ the PT profile becomes more monotonically increasing with pressure compared to a smaller value as presented in \citet{Barrado2023}. The retrievals are run with N = 1000 live points in constant efficiency mode and sampling efficiency of 0.05.

    \subsection{Self-consistent grid models}
    To compare the results from the atmospheric retrievals to self-consistently calculated model grids, we fit the data to grid models using the python package \texttt{species} \citep{Stolker2020} together with the sampling implementation \texttt{pyMulitnest}. The fitted grid models are: ATMO++ (no PH3) \citep{Meisner2023, Leggett2021}, six models by Lacy and Burrows: cloudy (thick clouds), cloudy (thin clouds) and clear either in chemical equilibrium or disequilibrium \citep{Lacy2023}, and the Sonora Elf Owl model \citep{Mukherjee2024}. 
    The grid models can be distinguished in four groups accounting for either cloudy or non-cloudy atmospheres and equilibrium or disequilibrium chemistry and their combinations. We fit one clear and equilibrium chemistry model: Lacy EQ Clear; three clear and non-equilibrium chemistry models: Lacy NEQ Clear, Sonora Elf Owl, ATMO++ (no PH3); two cloudy and equilibrium chemistry models: Lacy EQ cloudy (thick and thin); and two cloudy and disequilibrium chemistry models: Lacy NEQ cloudy (thick and thin). In all subsequent figures, clear and equilibrium chemistry is presented in brown colors, clear and disequilibrium chemistry in reddish colors, cloudy and equilibrium in greenish colors and cloudy and disequilibrium in bluish colors. 
    
    The cloudy model by \citet{Lacy2023} is available in two different cloud layer thicknesses and they assume a grain size of 10 $\mu$m. The thick cloudy Lacy models correspond to the AEE10 grid spectra and the thin cloudy models to the E10 spectra. For the non-equilibrium models the mixing parameter $K_{zz}$ is set to 10$^6$ $\mathrm{cm^2/s}$. Using this model grid we fit the effective temperature, the logarithmic gravity and the metallicity. 
    
    We compare the data to the Sonora Elf Owl grid. This model probes brown dwarfs of all spectral types ranging down to temperatures of 275 K. The models fit simultaneously for the temperature, gravity, metallicity, C/O ratio as well as the eddy diffusion coefficient $K_{zz}$. 
     
    We fitted the ATMO++ model for the effective temperature and the gravity \citep{Leggett2021,Meisner2023}. This model removes PH$_3$ as an opacity and includes an adaption to the slope of the PT structure in lower atmospheres as discussed in \citet{Tremblin2015,Leggett2021} to account for the cooling effect of fingering convection. The strength of the effect is parameterized using the factor $\gamma$, which is set to 1.3 in the used model.
    
    For the grid model fits we used the same binned data as for the retrieval analysis as mentioned in Section \ref{met:atmoret}. For the grid model comparisons, we add photometry data from Spitzer, HST and WISE as they are presented in Table \ref{table:photom} as presented in the 'Y dwarf compendium'\footnote{https://sites.google.com/view/ydwarfcompendium/y-dwarfs/wise-0855-0714}.

    \begin{table}
    \caption{The photometry used for the self-consistent grid model fits. References correspond to (1) \citet{Luhman2016}, (2) \citet{Schneider2016}, (3) \citet{Kirkpatrick2019}, (4) \citet{Wright2014} and (5) \citet{Leggett2017}. }             
    \label{table:photom}      
    \centering                          
    \begin{tabular}{l l l }        
    \hline\hline                 
    Filter &  Photometry [mag] & Reference \\    
    \hline                        
       HST/WFC3\_IR F110W     & 26.41 $\pm$ 0.27 & 1\\  
       HST/WFC3\_IR F160W     & 23.86 $\pm$ 0.03 & 2\\
       HST/WFC3\_IR F105W     & 27.33 $\pm$ 0.19 & 1\\
       HST/WFC3\_IR F127M     & 24.52 $\pm$ 0.12 & 1\\
                             & 24.49 $\pm$ 0.11 & 1\\
                             & 24.36 $\pm$ 0.09 & 1\\
       HST/ACS\_WFC F850LP    & 26.85 $\pm$ 0.44 & 1\\ \hline
       Spitzer/IRAC I1       & 17.470 $\pm$ 0.066 & 3\\ 
       Spitzer/IRAC I2       & 13.923 $\pm$ 0.016 & 3\\ \hline
       WISE/W1               & 17.819 $\pm$ 0.327 & 4\\
       WISE/W2               & 14.016 $\pm$ 0.048 & 4\\ 
       WISE/W3               & 11.9 $\pm$ 0.3 & 5\\ 
    \hline                                   
    \end{tabular}
    \end{table}
    
    In addition, to the above mentioned parameters we fit also for the object's radius. Using the values for the radius R and the surface gravity g, we calculate the mass using $M=gR^2/G$, where G is the gravitational constant \citep{Petrus2023}. We obtain luminosity estimates by integrating the flux of the best fit models using the built-in function in the package \texttt{species}.

    \begin{figure}[h]
        \centering
        \includegraphics[width=1\hsize]{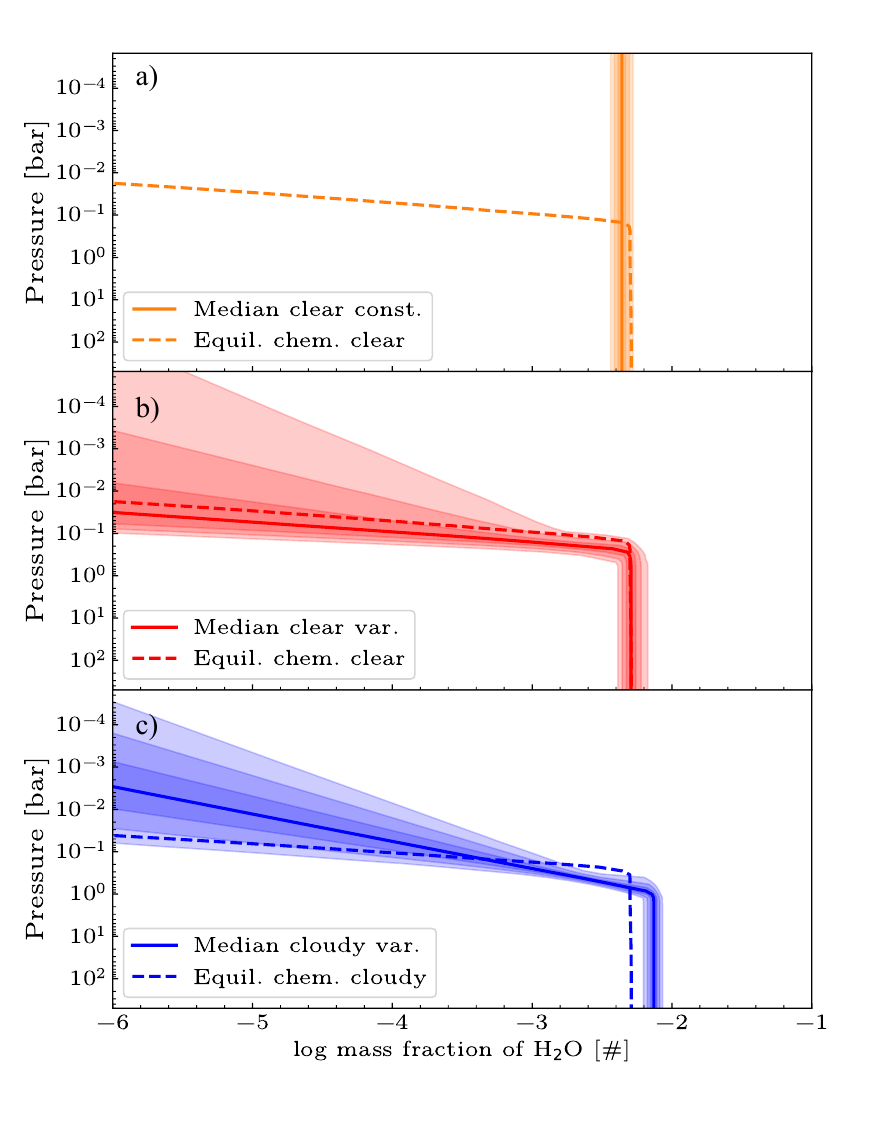}
        \caption{Logarithmic mass mixing ratios varying with pressure for the three shown retrieval results. In a) we show the retrieved median clear and constant water profile in a solid orange line and in the dashed line the equilibrium chemistry prediction based on the retrieved PT structure. Subpanel b) shows the analog for the clear retrieval with variable water profile in red and subpanel c) for the cloudy and variable water profile retrieval in blue. One to three sigma envelopes are shown around the retrieved profiles. The profiles where parameterized using Eq. \ref{eq:waterparam}. }
        \label{fig:abund_clouds}
    \label{fig:isotop}
    \end{figure}

    \begin{figure*}[h] 
    \centering
    \includegraphics[width=1\textwidth]{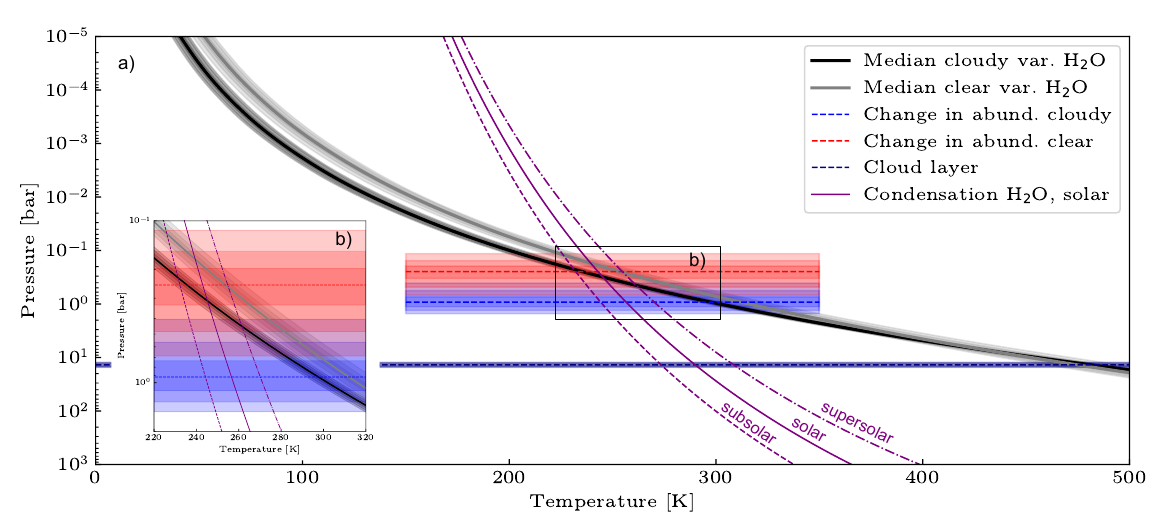}
    \caption{In panel a) pressure temperature lines for the clear retrieval in grey and the cloudy retrieval in black cross the 
    water condensation line for sub-, super- and solar metallicities \citep{Lodders2002} shown in purple. In a dashed red line we present the freely chosen atmospheric layer, where the water abundances is decreased for the clear retrieval and in blue for the cloudy retrieval. We show one to three sigma envelopes around the median value for the change in water abundance. For the clear retrieval the crossing between the condensation line and the PT profile happens at the pressure level chosen freely by the retrieval for the reduction in water abundance indicating that water is likely to condensate out at this height. In panel b) we show an inset to better visualize the crossing area. The dark blue dashed line corresponds to the cloud layer constrained by the retrieval.} 
    \label{fig:h2ointersec}
    \end{figure*}

    \begin{figure*}[h]
        \centering
        \includegraphics[width=1\hsize]{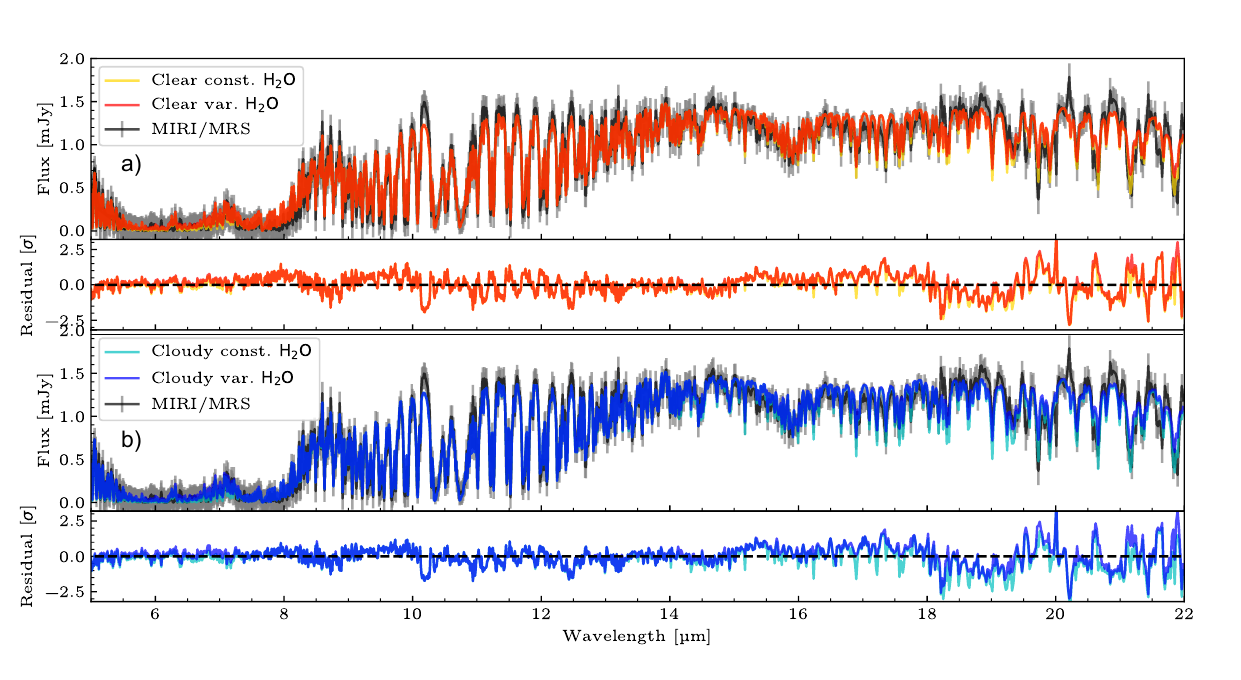}
        \caption{The effect of the water depletion on the spectrum. The upper plot a) compare the clear model with the constant water profile in yellow and the variable in red to the data in black. The lower plot b) compares the same for the cloudy model with the constant profile in cyan and variable in blue. In both panels, we use except for the profile parameters the retrieved values from the variable retrieval as well for the constant model.}
        \label{fig:waterdepl}
    \end{figure*}

    \begin{figure*}[h] 
    \centering
    \includegraphics[width=1\textwidth]{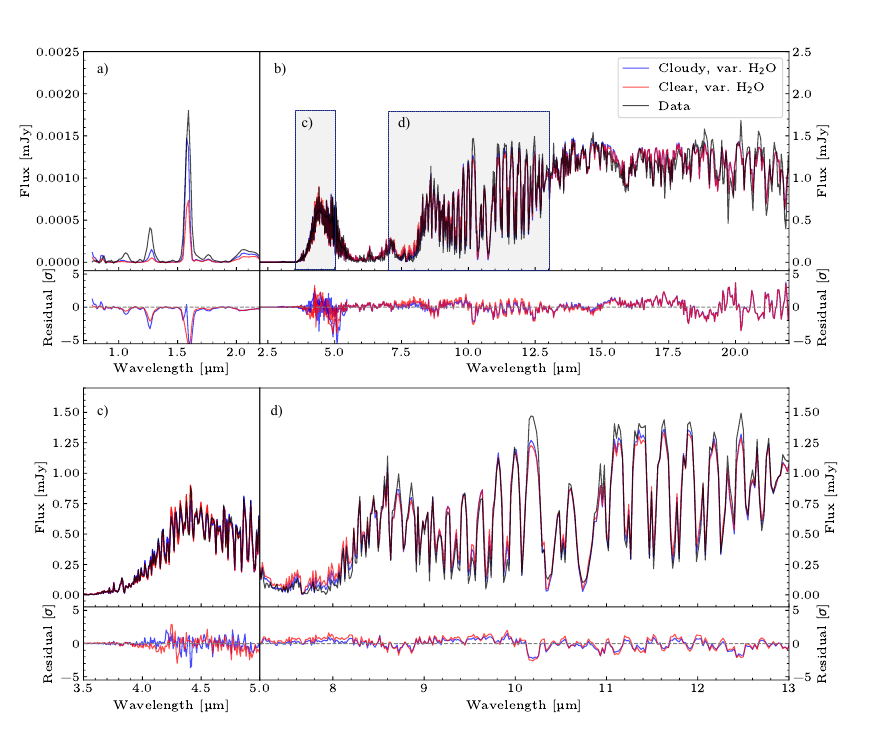}
    \caption{Best fit retrieval spectra of the cloudy in blue and the clear retrieval in red. Both models include a variable water profile and are compared to the data in black. Panel a) shows the data and retrievals at the wavelength range from 0.8 to 2.2 µm, panel b) at the wavelength from 2.2 µm to 22 µm, c) presents the inset from 3.5 to 5.4 µm and d) from 7.2 to 13 µm. The latter two panels are included in panel b) where their locations are highlighted. In every panel we show the flux in the upper part and the corresponding residual between retrieval and data in the lower plot. For better visibility the presented models and data are re-binned to a resolution of R = 500 for the NIRSpec/G395M and MIRI/MRS and R = 100 for NIRSpec/PRISM wavelengths.}%
    \label{fig:cloud_ret}
    \end{figure*}

    \begin{figure*}[h]
    \centering
    \includegraphics[width=1\textwidth]{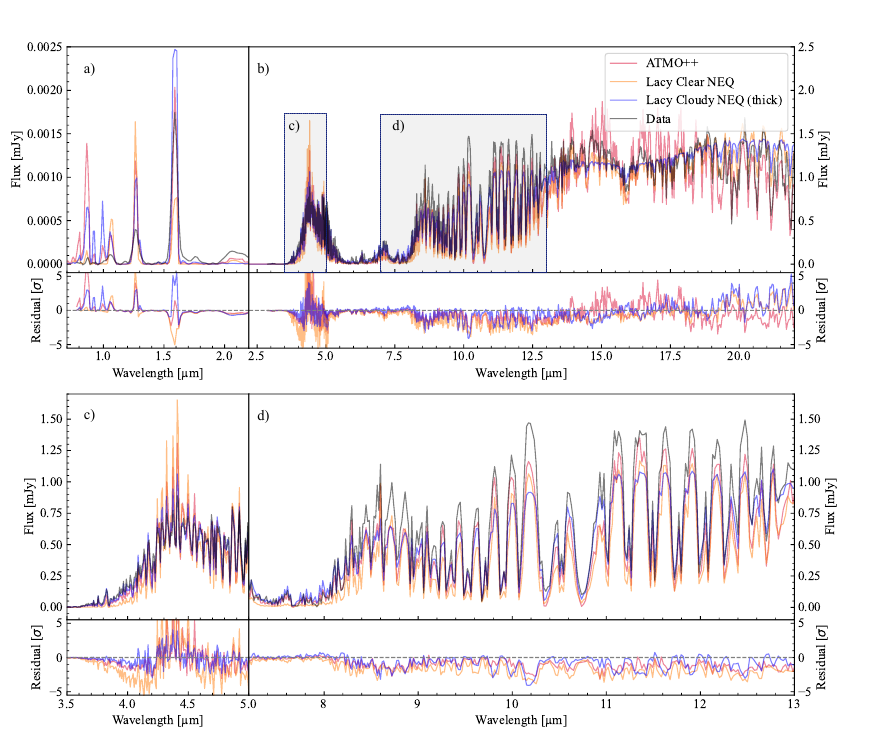}
    \caption{Best fit grid model spectra for the clear ATMO++ in red, the thick cloudy non-equilibrium model by \citet{Lacy2023} in blue and and the clear non-equilibrium in orange. Subplot a) shows an inset of the PRISM data, which has a significant lower amount of flux. Subfigure b) gives the NIRSpec/G395M and MIRI/MRS data set. Subplot c) and d) show insets in panel b) on the NIRspec/G395M region and the prominent NH$_3$ absorption feature. The corresponding $\chi^2$ values are 2.57, 1.97 and 5.66 for ATMO++, the Lacy cloudy (thick) and clear non-equilibrium model fits. As for the retrievals and for better visibility we re-bined the models and data to a resolution R = 500 for the NIRSpec/G395M and MIRI/MRS and kept NIRSpec/PRISM at R = 100.}%
    \label{fig:cloud_self}
    \end{figure*}
    
    \section{Results} \label{results}
    In the first part, we will focus on the water depletion in WISE~0855's atmosphere. Subsequently, we will present the self-consistent models in particular with respect to cloudy compared to clear atmospheres and chemical (dis)equilibrium. Finally, we will present the ammonia isotopologue detection and the estimate for the bolometric luminosity. 
    
    \subsection{Water depletion from atmospheric retrievals} 
    \label{res:conden} 
    Atmospheric retrievals can infer information about the composition at different pressure levels. In this retrieval setup we vary the water profile using the parameterization presented in Eq. \ref{eq:waterparam} and compare it to the retrieval results from a constant water abundance with height. The pressure level where the retrieval may change the abundance in the atmosphere can be chosen freely. We want to understand whether the retrieval changes the abundance and if so, if the retrieved value can be linked to condensation. 

    In Figure \ref{fig:abund_clouds} we present the fitted mass mixing ratios of water for the three retrievals with cloudy and variable profile in blue, with clear and variable profile in red and with clear and constant water profile in orange. In all cases the shaded areas correspond to one to three sigma variations in the profiles. The chemical equilibrium values are calculated with \texttt{easyCHEM}, an open-source\footnote{\url{https://easychem.readthedocs.io/en/latest/}} Gibbs free energy minimizer to calculate chemical equilibrium compositions presented in \citet{Molliere2017}. A Table interpolating \texttt{easyCHEM} results (also including the condensation of water) is available in \texttt{petitRADTRANS}, and we plot the H$_2$O steam abundance for solar metallicity ([Fe/H] = 0) and C/O ratio (C/O = 0.55). The blue dashed line uses the temperature pressure profile from the cloudy retrieval, the red dashed line from the clear retrieval and the orange dahsed line from the clear and constant retrieval.
    
    In the clear case the bend of the equilibrium profile is inside the one sigma range of the fitted profiles. The initial logarithmic water abundance is well constrained and in the two clear cases it is comparable with the equilibrium chemistry prediction of $-2.29$ with logarithmic values of $-2.29 \pm 0.03$ for the clear and variable profile retrieval and $-2.36 \pm 0.03$ for the clear and constant profile retrieval. We find an initial logarithmic water abundance of $-2.13 \pm 0.02$ for the cloudy retrieval, which is slightly larger compared to the chemical equilibrium expectation. The decrease in abundance of the median profile of the clear retrieval fits the chemical equilibrium expectations well even though the variation is large. The cloudy retrieval decreases less with higher layers compared to the chemical equilibrium prediction. We obtain a pressure where the retrieval changes the water abundance at 10$^{-0.04 \pm 0.10}$ bar for the cloudy retrieval and 10$^{-0.60 \pm 0.12}$ bar for the clear retrieval. The slope for the change in the variable water profile is constrained as $\alpha$ = 1.5 $\pm$ 0.5 for the cloudy and 4.3 $\pm$ 2.3 for the clear case.

    Figure \ref{fig:h2ointersec} shows the pressure-temperature structures of the retrievals in black and grey lines for the cloudy and clear retrievals, respectively. The profiles show the median of 1000 randomly chosen profiles from the posterior distribution with one to three sigmas variation of the in shaded regions. The retrieved PT structure of the cloudy retrieval is shifted to higher pressures. The purple lines correspond to the condensation lines of water for either sub-, super- or solar metallicities ([Fe/H]=-0.5, 0.5, 0) from \citet{Lodders2002}. The dashed line in red shows the pressure level freely chosen by the retrieval, where the water abundance is reduced compared to the initial water abundance. The location of the constrained cloud layer by the cloudy retrieval is shown in dark blue. 
    
    Interestingly, for the clear retrieval the PT profile crosses the water condensation line at the height where also the change in water abundance happens. As the water abundance above the corresponding pressure level decreases (as shown in Figure \ref{fig:abund_clouds}) and we allow the retrieval to choose the pressure parameter and the reduction in water abundance freely, this provides indication for water condensation in WISE~0855's atmosphere. The cloudy retrieval reduces the abundance of water as well, however slightly lower in the atmosphere than the crossing between the condensation line and the PT profile. The additionally constrained cloud parameters might affect the height of change in water abundance.

    In Figure \ref{fig:waterdepl} we present the resulting spectra from the forward model used in the retrievals with either a constant or variable water profile to present the effect of the water reduction on the spectrum. We use retrieval results from the variable water profiles as inputs to the models either with a constant or with a variable water profile. We see in both the clear and cloudy case that the spectrum with a variable water profile fits the depth of the water absorption bands better from about 14 µm to 22 µm compared to the one with a constant profile. The effect is slightly stronger in the cloudy case, as here the location from which the water abundance decreases is  even lower in the atmosphere. In addition, we can see a slight improvement of the fit between 6 and 7 µm, corresponding to the water absorption band.

    The variable water profile fits significantly better compared to the constant water profile. We find a Bayes factor between the clear and constant water profile retrieval (ln(Z) = 20461) and the clear retrieval with variable water profile (ln(Z) = 20468) of ln(B) = 7 corresponding to a 4.2 $\sigma$ significance. For the cloudy case, between the retrieval with a constant (ln(Z) = 20568) and the variable water abundance (ln(Z) = 20602) we find a Bayes factor of ln(B) = 34 corresponding to a 8.3 $\sigma$ significance \citep[following the method of][to convert $\Delta {\rm ln}Z$ to detection significances]{Benneke2013}.

    \subsection{Clear vs. cloudy atmospheric retrievals}
    \label{res:clouds}
    Figure \ref{fig:cloud_ret} shows the best fit models of the cloudy and clear retrievals for a variable H$_2$O profile. 
    Panel a) shows the fit of the NIRSpec/PRISM data, where we find a significantly better agreement of the 1.6 µm flux peak for the cloudy model compared to the clear model. Both models underestimate the 1.3 µm flux and the clear model more than the cloudy one. In panel b) we present the fit from 2.2 to 22 µm showing the two insets presented in c) and d). The residuals become larger beyond 15 µm for both models.
    In panel c) we show the inset to the NIRSpec/G395M wavelength, agreeing well with the data. The residuals are slightly larger between 4.3 and 4.5 µm in the cloudy model compared to the clear model. The methane and ammonia features between 7 to 13 µm are presented in panel d). Between 7 and 12.5 µm, we see a better fit of the cloudy model compared to the clear one of the order of 2 $\sigma$ in residuals. Here, the clear model underestimates the flux in the peaks and overestimates the flux in the valleys of the absorption lines by about 1.5$\times 10^{-4}$ Jy. The cloudy model fits better and the residuals are smaller while maintaining the residual shape. Between 7 and 8.5 µm the cloudy retrieval models the CH$_4$ band better compared to the clear retrieval, while the clear model shows an offset compared to the data. The abundances found by the retrievals are presented in Table \ref{table:posteriors} and their posterior distributions in Figure \ref{fig:corner_abund} in the appendix.
    
    The cloudy retrieval constrains the cloud parameter, as presented in the corner plot in the appendix in Figure \ref{fig:corner_cloud}. We find a mean cloud particle logarithmic mass fraction of -3.11 $\pm$ 0.30 at the base, a ${\rm log}$ cloud base pressure at 1.14 $\pm$ 0.03, a $\rm \sigma_{lnorm}$ of 1.39 $\pm^{+0.37}_{-0.22}$, a logarithmic eddy diffusion factor $\rm K_{zz}$ of 1.3 $\pm^{+0.1}_{-0.2}$, a sedimentation parameter $\rm f_{\rm sed} = 10.2^{+0.5}_{-0.7}$. 
    
    When comparing the global logarithmic evidence of the two retrievals with variable water abundance using the logarithmic Bayes factor ln(B), we find that the cloudy model (ln(Z) = 20602) fits significantly better when compared to the clear fit (ln(Z) = 20468) by ln(B) = 134, corresponding to 16.6 sigma using \citet{Benneke2013}. 

    The cloudy retrieval constrains the cloud parameter and the Bayes factor indicates evidence for the cloudy model. However, the cloud layer is set deep in the atmosphere at temperatures of about $\sim$500 K, where it is not possible to form water ice clouds and where they would not be stable. Thus, even though the Bayes factor suggests to include clouds, we do not detect water ice clouds. However, this cloud might still be real and accounting for another potentially phosphorous-bearing species condensing out in this depth. 

    \subsection{Clear vs. cloudy self-consistent models}
    To complement the retrieval analysis we fit our data with more physics-informed models such as the radiative-convective equilibrium models by \citet{Lacy2023}, ATMO++ without PH$_3$\citep{Leggett2021,Meisner2023} and Sonora Elf Owl \citep{Mukherjee2024}. In Figure \ref{fig:cloud_self} we compare the three best-fit results of the clear, the cloudy (thick) models by \citet{Lacy2023} and ATMO++. The resulting reduced $\chi^2$ values are 5.66, 1.97 and 2.57, respectively. By comparing the best fit models to the data we see the clear model by \citet{Lacy2023} shows an overshoot in flux at around 4.3 µm as presented in Figure \ref{fig:cloud_self} c) and a constant underestimation in flux between 7.5 and 22 µm. The cloudy model explains the 4.3 µm better compared to the clear ones. The ammonia feature given in Figure \ref{fig:cloud_self} d) is best explained by the ATMO++ model, however still underestimating the flux. The ATMO++ model fits well the 1.6 µm peak and the NIRSpec/PRISM wavelengths. A list of reduced $\chi^2$ values is presented in Table \ref{table:goodnessfit_cons}.

    \begin{figure}[h]
        \centering
        \includegraphics[width=1\hsize]{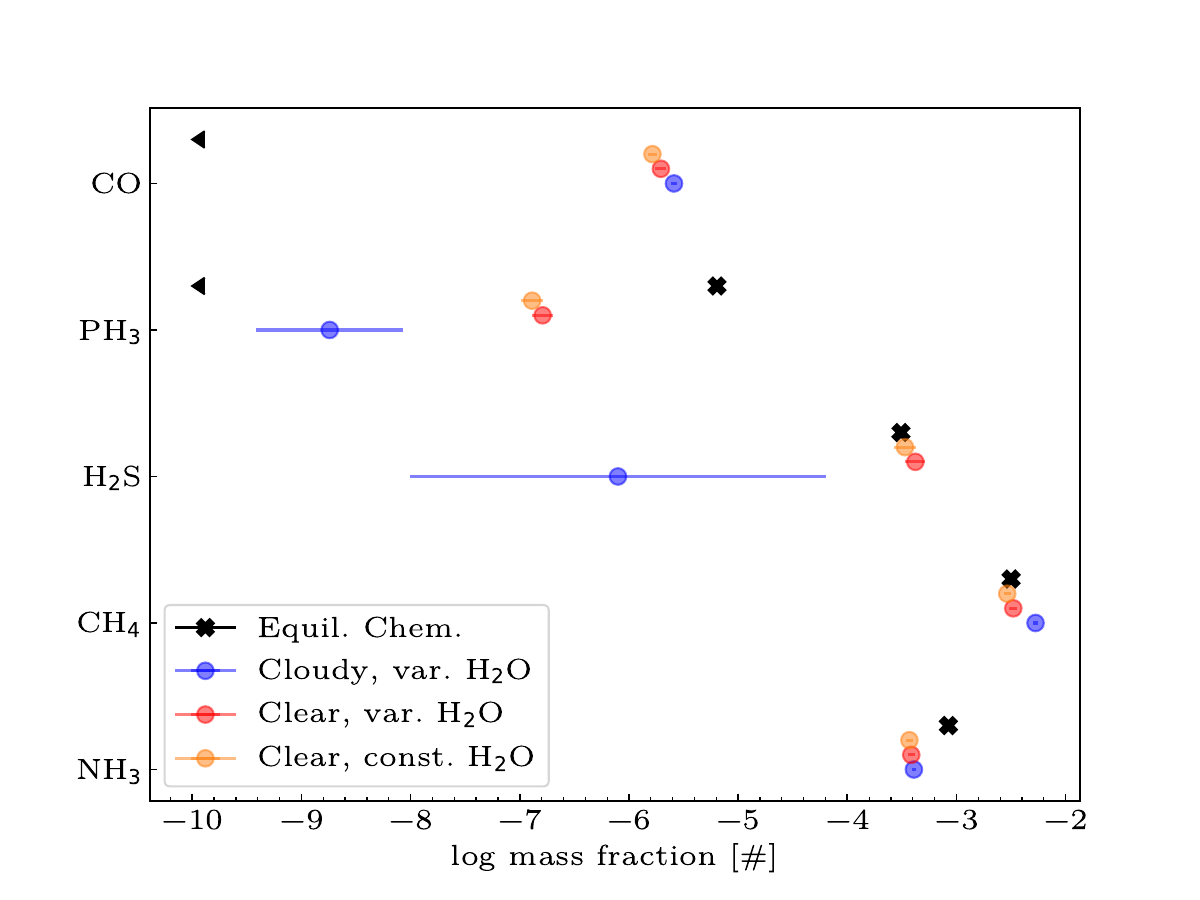}
        \caption{Retrieved logarithmic mass mixing ratios in comparison with chemical equilibrium values for the species: CO, PH$_3$, H$_2$S, CH$_4$, NH$_3$. Black crosses show the expected values for equilibrium chemistry calculated for the cloudy and variable water profile retrieval. Black triangles correspond to equilibrium values smaller than -10. For $\mathrm{PH_3}$ we obtain in the lower atmosphere the crossed value and in the upper atmosphere a value lower than the shown x-axis. We depict in blue the retrieved values from the cloudy and variable water profile, in red from the clear and variable water profile and in  orange the clear and constant water profile.}
        \label{fig:abund_diseq}
    \label{fig:isotop}
    \end{figure}

    \begin{figure*}[h]
    \centering
    \includegraphics[width=1\textwidth]{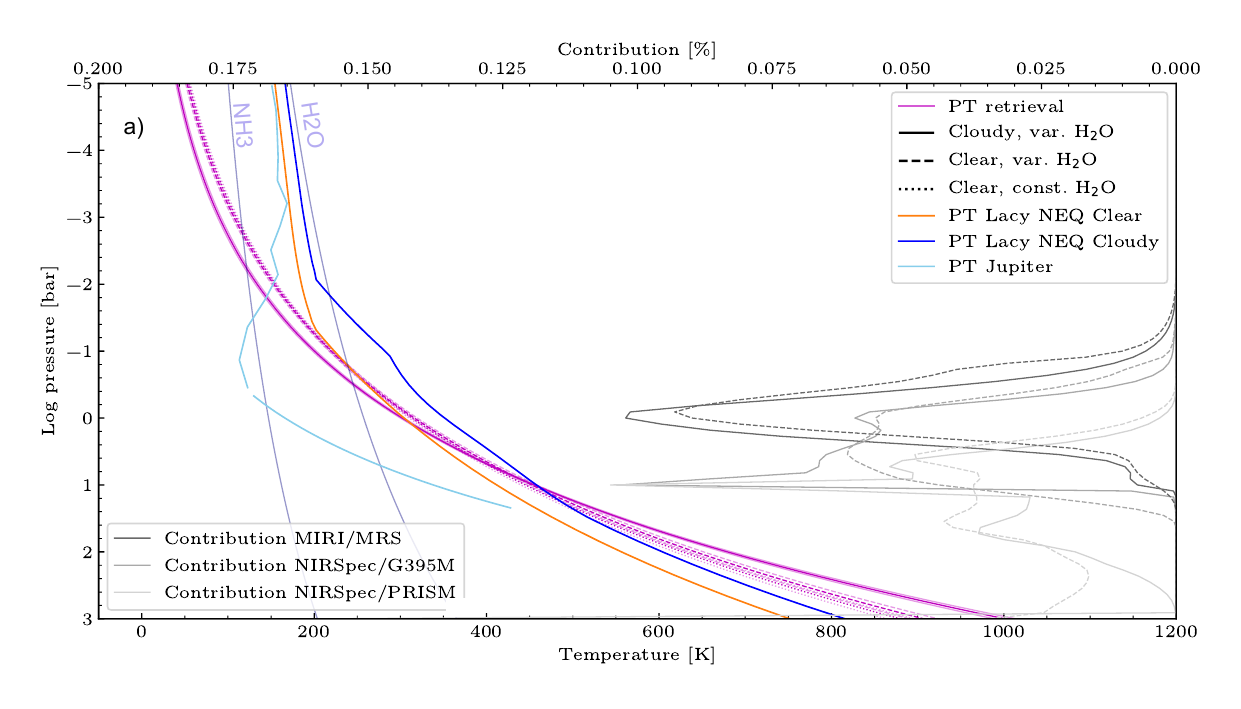}
    \caption{Pressure-Temperature (PT) profiles of the retrievals in magenta compared to the $\mathrm{H_2O}$ and $\mathrm{NH_3}$ condensation line in violet and the contribution functions of MIRI/MRS, NIRSpec/G395M and NIRSpec/PRISM in grey. Solid lines correspond to the cloudy, variable water profile, dashed lines to the clear and variable water profile and dotted lines to the clear and constant water profile retrievals.  We show the mean PT profiles of a 1000 randomly drawn profiles from the posterior distribution. Fainter lines correspond to plus minus one standard deviation from the distribution of profiles. In light blue we present the measured and interpolated PT profiles of Jupiter \citep{Seiff1998}. In blue we compare the pressure temperature profiles at 250K effective temperature of the Lacy NEQ Cloudy (thick) model for subsolar metallicity with the Lacy NEQ Clear model for supersolar metallicity in orange \citep{Lacy2023}. }%
    \label{fig:retrieval_pt}
    \end{figure*}

\subsection{Chemical disequilibrium in retrieved abundances} \label{res:diseqchem}
    
    The composition of the atmosphere at a certain pressure and temperature level gives hints about the dynamics in the atmosphere. From chemical equilibrium calculations we can derive what atmospheric composition we would expect if no mixing were present. If there is a difference in the observed composition compared to the expected one, this indicates that processes leading to non-equilibrium are present.
    
    In Figure \ref{fig:abund_diseq} we present the abundances of the trace gases as logarithmic mass mixing ratios resulting from the retrieval analysis for the species: $\mathrm{CH_4}$, $\mathrm{NH_3}$, $\mathrm{CO}$, $\mathrm{H_2S}$, $\mathrm{PH_3}$. We compare the retrieved values for all retrievals with the chemical equilibrium calculation presented in \citet{Molliere2017}. 
    
    $\mathrm{NH_3}$ shows slightly smaller values than expected from chemical equilibrium and is very well constrained by the retrievals. We find values of $-3.39 \pm 0.02$ for the cloudy retrieval, $-3.41 \pm 0.03$ for the clear and variable water profile retrieval and $-3.43 \pm 0.03$ for the clear and constant water profile retrieval. The equilibrium value is $-3.07$.
    We see well constrained values for CO, even though from chemical equilibrium calculations it is not expected to be present. CO has been previously detected by \citet{Miles2020}. Here, we find values of $-5.59 \pm 0.03$ for the cloudy retrieval, $-5.71 \pm 0.05$ for the clear and variable water profile retrieval and $-5.78 \pm 0.04$ for the clear and constant water profile retrieval. This is a strong indication for chemical disequilibrium as in the observed height, CO gets converted to CH$_4$ via chemical reactions. 
    $\mathrm{CH_4}$ is well constrained and shows similar values compared to what is expected from chemical equilibrium predicting a value of of $-2.50$. For the cloudy retrieval we find $-2.28 \pm 0.02$, for the clear and variable retrieval $-2.48 \pm 0.04$ and for the clear and constant water profile retrieval $-2.54 \pm 0.03$. 
    For PH$_3$ the cross marks the equilibrium chemistry abundance for pressures above 15 bar and the arrow for pressures below 15 bar as PH$_3$ is only expected in the deep atmosphere when in chemical equilibrium. $\mathrm{PH_3}$ seems to be constrained in the clear case with constant water profile to $-6.88^{+0.09}_{-0.10}$ and clear and variable water profile to $-6.79^{+0.09}_{-0.10}$. However in the cloudy case we find $\mathrm{PH_3}$ to be fairly unconstrained with a value of $-8.75 \pm 0.75$. $\mathrm{PH_3}$ has not been detected in WISE~0855 so far \citep{Luhman2024,Morley2018,Skemer2016} compatible with the cloudy case. Comparing the spectrum with the opacity spectra shown in Figure \ref{fig:opacities} in the appendix, we do not see the strong features of PH$_3$ at around 4.3 µm in the spectrum. This leads to the conclusion that the constrained values are compensating in the clear case for a mismatch in the flux. We find a similar effect with regard to $\mathrm{H_2S}$. For the clear and constant retrieval we estimate $-3.46^{+0.09}_{-0.10}$, for the clear and variable profile $-3.37^{+0.09}_{-0.10}$ and for the cloudy retrieval $-5.83^{+1.83}_{-2.58}$. The clear retrievals are constrained while the cloudy is unconstrained with a slight preference towards larger values. The equilibrium chemistry calculations predict a value of $-3.51$. 

    In Figure \ref{fig:selfcons_chem_cloudy} in the appendix we present the grid model fits for the chemical disequilibrium Sonora Elf Owl model as well as for the Lacy clear and equilibrium model. The Sonora Elf Owl leads to a $\chi^2$ value of 6.77 and the clear Lacy equilibrium model to the highest $\chi^2$ in this comparison of 12.72. In the Figure \ref{fig:selfcons_chem_clear} in the appendix we present the other cloudy models by \citet{Lacy2023} for either chemical equilibrium or disequilibrium and two different cloud heights. They result in a $\chi^2$ of 2.75, 3.48, 1.97 and 2.48 for the thick and thin cloudy equilibrium chemistry and the thick and thin cloudy disequilibrium chemistry.

    \subsection{Grid model comparison} \label{res:modcomp}   
    We compare the data of WISE~0855 to the eight different self-consistent grid models previously discussed in Section \ref{res:clouds} and \ref{res:diseqchem}. We compare the resulting outputs in Figure \ref{fig:selfcons_comp} and in Table \ref{table:gridmodres} both in the appendix. 
     
    Across the used grid models, the median effective temperature is $\mathrm{T_{eff}}$ = 250.4 $\pm$ 6.2 K. The cloudy Lacy models consistently result in an effective temperature of 250 K except of the clear Lacy equilibrium model resulting in a temperature of 261.4 $\pm$ 0.6 K and the non-equilibrium cloudy (thin) model in 248.8 $\pm$ 0.5 K being the coldest estimate for WISE~0855 stated yet. The clear ATMO++ model result in significantly higher temperatures of 297.4$ \pm$ 0.7 K for the one including and neglecting PH$_3$ respectively. The Sonora Elf Owl model also reaches the lower temperature limit being  $\mathrm{T_{eff}}$ = 275.0 $\pm$ 0.1 K. For the latter we obtain a value for the mixing parameter log(K$_{zz}$) = 2.25 $\pm$ 0.04 cm$^2$/s and slightly sub-solar C/O ratio of 0.51 $\pm$ 0.01. However, as WISE~0855 might be even colder than the lower limit of the temperature range of this grid, it is generally difficult to interpret the results from this model.  
    
    The gravity estimates are consistently reaching the lower limit of the grids being log(g) = 3.5 cm/s$^2$ for the Lacy models and 3.23 cm/s$^2$ for the Sonora Elf Owl. They hit the lower boundary of the grid model and thus result in very small errors smaller than 0.001 cm/s$^2$. The ATMO++ model reaches a larger gravity of 4.2 cm/s$^2$. In general, gravities estimated by grid models are smaller compared to the estimate from retrieval analysis, which result in gravities of 4.7 cm/s$^2$ for the clear models and 4.9 cm/s$^2$ for the cloudy model - this is more than ten times larger compared to the self-consistent model estimate. Literature values range from log(g) = 3.5 to 4.5 cm/s$^2$ for a cold brown dwarf like WISE~0855 based on the Sonora Bobcat evolutionary models \citep{Miles2020}. As the grid fits hit often the lower boundary, we fitted the grid models constraining the surface gravity to log(g) = 4.0 - 5.0 to probe for other global minima when neglecting the lower gravity bound. The resulting grid parameter are presented in Table \ref{table:gridmodres} in the appendix. Except for ATMO++ and Lacy non-equilibrium clear, the models still go to the lower bound of the surface gravity.
    
    The radius is consistently estimated at around 1 $\RJ$. The Lacy models predict a slightly larger than Jupiter radius of about 1.15 - 1.40 $\RJ$ except for the clear equilibrium model reaching lower radii of 0.94 $\RJ$. ATMO++ reaches the lowest radius of 0.79 $\RJ$ and Sonora Elf Owl slightly larger values of 0.98 $\RJ$. 

    The mass is calculated from the radius and gravity estimates. For the Lacy models except for the clear and equilibrium one we estimate masses of about 2 $\MJ$. Cloudy models predict a higher mass compared to clear models. For the ATMO++ fit we estimate a larger mass of about 3.7 $\MJ$, while Sonora Elf Owl reaches an unrealistically small mass of 0.33 $\MJ$.
    
    The cloudy Lacy models result in subsolar metallicities ranging from [Fe/H] = -0.2 to -0.4. The clear Lacy models and Sonora Elf Owl shows a supersolar metallicites of [Fe/H] = 0.4 - 0.5. The metallicity of the ATMO++ grid is fixed at solar metallicity.

    By integrating the best fit spectra of the self-consistent grid models, we obtain estimates for the luminosity compared to solar. The values are consistent with a median value of -7.294 $\pm$ 0.023 in logarithmic scale. Cloudy models tend to result in slightly lower values compared to clear models, except of the ATMO++ models reaching similar values compared to the cloudy Lacy models. The values are presented in Table \ref{table:gridmodres_lum} in the appendix.

    The reduced $\chi^2$ values for the various models are presented in Table \ref{table:goodnessfit_cons}. The two smallest reduced $\chi^2$ are given by the Lacy non-equilibrium model with thick clouds, followed by the Lacy non-equilibrium model with thin clouds and the ATMO++ model. The largest reduced $\chi^2$ results from the Lacy clear and equilibrium model. 
    
    \begin{table}
    \caption{Reduced $\chi^2$ values for the self-consistent models ordered from the lowest values to the highest. The values were calculated including the $10^b$ error from the clear retrieval presented in Table \ref{table:posteriors}}             
    \label{table:goodnessfit_cons}      
    \centering                          
    \begin{tabular}{l | l}        
    \hline\hline                 
    Model & red $\chi^2$   \\    
    \hline                        
       Lacy non-eq. cloudy (thick)  & 1.97    \\
       Lacy non-eq. cloudy (thin)   & 2.48    \\
       ATMO++                       & 2.57    \\
       Lacy eq. cloudy (thick)      & 2.75    \\
       Lacy eq. cloudy (thin)       & 3.48    \\
       Lacy non-eq. clear           & 5.66    \\ 
       Sonora Elf Owl               & 6.77    \\
       Lacy eq. clear               & 12.72   \\
       \hline                         
    \end{tabular}
    \end{table}

    \begin{figure*}[h]
        \centering
        \includegraphics[width=1\textwidth]{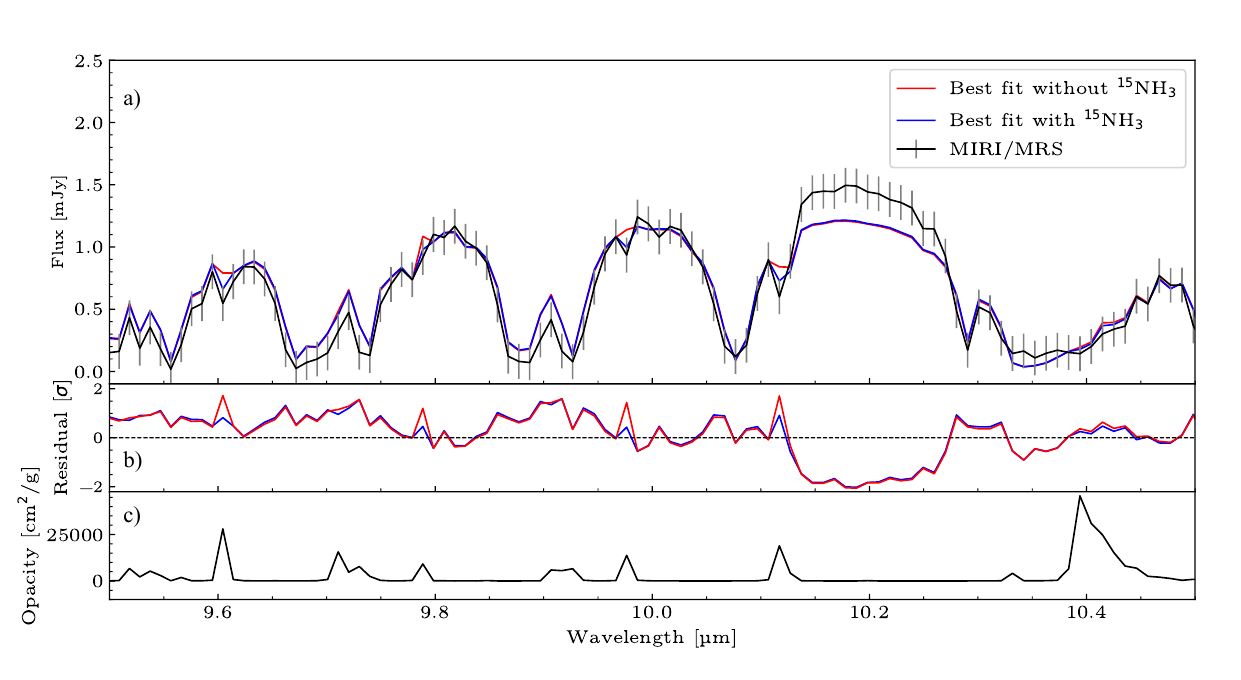}
        \caption{We show the best fit spectrum with $^{15}$NH$_3$ included in blue, the best fit removing the $^{15}$NH$_3$ opacity in red and in black the data with error bars. The top right panel b) shows the $^{15}$NH$_3$ feature observed by the MIRI/MRS spectrum of WISE~0855 with an inset shown in panel a). The middle plot d) shows the residuals of the fit with and without $^{15}$NH$_3$ in blue and red respectively. The lowest plot f) shows the $^{15}$NH$_3$ opacity used for the retrieval at 277 K and for the pressures of the clear retrieval with varying H$_2$O profile. Insets of the middle and lower plots are shown in panels c) and e)}
    \label{fig:isotop}
    \end{figure*}
    
    \begin{figure}[h]
        \centering
        \includegraphics[width=1\hsize]{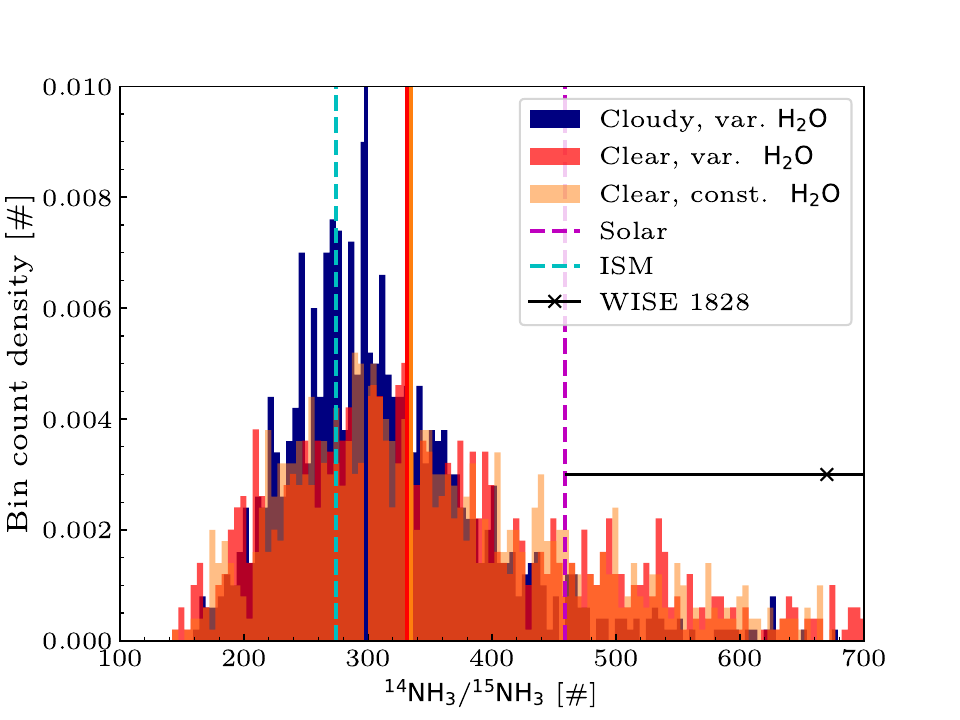}
        \caption{Distribution of the $\mathrm{^{14}NH_3}$/$\mathrm{^{15}NH_3}$ ratios for each retrieval with the solid lines marking the corresponding median value. The cyan colored line represents the value for ISM and the magenta colored line the value for the Sun. The value presented by \citet{Barrado2023} for the brown dwarf WISE~1828 is shown in black and the line indicating the uncertainty.}
    \label{fig:ammoniadistr}
    \end{figure}

    \subsection{Atmospheric structure} \label{res:atmos}
    The atmospheric structure returned by the retrievals and Lacy grid models is presented by the PT profiles in Figure \ref{fig:retrieval_pt}. We compare the retrieved PT profiles to the profiles of the best-fitting models by \citet{Lacy2023} as well as to the PT profile from Jupiter measured by the Galileo probe \citep{Seiff1998}. In addition, we present the contribution function, showing us which atmospheric layer the observed flux per instrument originates from.
    
    In general, the retrieved PT profiles are very similar to each other, however the clear structures seem slightly steeper compared to the cloudy one. Also the variation in the PT structures are very small especially between 0.3 to 30 bars where we have contributions from the data. All retrieved PT profiles cross the water condensation line in the area visible by MIRI/MRS as indicated by the contribution functions in grey. In the retrieval we introduced a regularization factor $\gamma$ which penalizes inversions in the PT structure \citep{Line2015}. This value is well constrained and we obtain values of 6.6 $^{+1.0}_{-0.9}$ $\times$ 10$^{-7}$ for the cloudy, 3.4 $^{+0.6}_{-0.5}$ $\times$ 10$^{-7}$ for the clear and variable water profile and 2.9 $^{+0.5}_{-0.4}$ $\times$ 10$^{-7}$ for the clear and constant water profile retrieval. As larger values correspond to larger constraints on the variability, we penalize the cloudy profile more than the clear ones. This might lead to the observed steepness difference in the two profiles. 
    
    We plot the PT structures of the Lacy NEQ clear and cloudy (thick) models at an effective temperature of 250 K and log(g) = 3.5 at supersolar and  subsolar metallicity in orange and blue respectively. In general, the Lacy profiles are slightly steeper in height compared to the retrieval outputs. Shortly after crossing the water condensation line with height the Lacy models become nearly isothermal along the condensation line while the retrieved PT profiles are not as steep. The cloudy Lacy PT shows a constant offset compared to the clear one. Higher metallicities shift the PT structure to lower pressures and larger gravities shift the PT structure to higher pressures \citep{Molliere2015,Fortney2018}. As the Lacy models find lower gravities compared to the retrievals, they will probe the atmosphere at lower pressures. The cloudy Lacy PT profile shows a similar behaviour as the retrievals in the lower part of the atmosphere until it becomes warmer compared to the clear profile at around 0.1 bar following the line of the water condensation. 
    
    We show the mean contribution function over the posterior distribution and wavelengths per instrument from the retrievals, showing the height in the atmosphere where the flux per wavelength in the retrieved spectrum originates from. In dark grey we present the contribution of MIRI/MRS, in medium grey of NIRSpec/G395M and in light grey of NIRSpec/PRISM. At longer wavelengths we are probing colder areas of the atmosphere corresponding to higher altitudes. With MIRI/MRS we are thus probing a region of about 0.3 to 10 bar, with NIRSpec/G395M about 1 to 30 bar and with NIRSpec/PRISM about 3 to 100 bar. The contribution of NIRSpec/G395M shows a strong peak at around 13 bar. This is the pressure level where the clouds are constrained in the cloudy retrieval. However, the clouds seem to be not completely opaque, as the retrieval shows contributions below the cloudy layer with NIRSpec/PRISM down to even 1000 bar. As we do not have many data points in the visible wavelength range, this estimate needs to be taken with caution. 

    Jupiter's PT profile is similar to the retrieved profile with a constant offset of about 150 K between about 0.3 and 30 bar, which is also the area where we probe the atmosphere with our data. The upper atmosphere for pressures lower than 0.3 bar the PT profile of Jupiter becomes nearly isothermal. We can see that at about the same height where Jupiter crosses the ammonia condensation line, WISE~0855 crosses the water condensation line for the retrieved PT structures and the clear Lacy model.

    \subsection{$\mathrm{^{15}NH_3}$ isotopologue abundance} \label{res:isotop}
    The resolution of MIRI/MRS of up to R $\sim$ 3'500 in the mid infrared allows us to search for isotopologues. We detect the isotopologue $\mathrm{^{15}NH_3}$ in WISE~0855's ammonia-rich atmosphere. Comparing the evidence of the atmospheric retrieval with and without $\mathrm{^{15}NH_3}$ as shown in Table \ref{table:goodnessfit_ret} results in a logarithmic Bayes factor ln(B) = 10 favouring the model including $\mathrm{^{15}NH_3}$ by 4.9 $\sigma$ when comparing to \citet{Benneke2013}. In Figure \ref{fig:isotop} we present the spectral differences in the models with and without $\mathrm{^{15}NH_3}$. We find a clear signature of $\mathrm{^{15}NH_3}$ absorption at 9.26, 9.61 µm, 9.97 µm and 10.11 µm. Here, the residuals of the retrieval without $\mathrm{^{15}NH_3}$ and the opacities match very precisely at the same wavelengths enforcing the detection. 

    From the mass mixing ratios we can calculate the value $\mathrm{^{14}NH_3}$/$\mathrm{^{15}NH_3}$. The distribution of the values is presented in Figure \ref{fig:ammoniadistr}. For the median over a sample of 1000 spectra from the posterior distribution, we obtain a value of $\mathrm{^{14}NH_3}$/$\mathrm{^{15}NH_3}$ = 299$^{+35}_{-29}$ for the cloudy and variable water profile retrieval, $\mathrm{^{14}NH_3}$/$\mathrm{^{15}NH_3}$ = 332$^{+63}_{-43}$ for the clear and variable water profile retrieval and $\mathrm{^{14}NH_3}$/$\mathrm{^{15}NH_3}$ = 335$^{+63}_{-46}$ for the clear and constant water profile retrieval. The uncertainty is given by the differences between the median and the 32nd and 68th percentile respectively of the mentioned distribution. The distribution of the values shows a non Gaussian shape with a tail towards larger values. Interestingly, the tail is more pronounced for clear retrievals compared to the cloudy retrieval, which seems to follow more a Gaussian distribution. Thus, the median value is slightly lower for the cloudy retrieval compared to the clear retrievals. In Figure \ref{fig:ammoniadistr} we further compare the ratio to values of the Interstellar Medium (ISM), the Sun, and the first detection in the cold brown dwarf WISE~1828 presented in \citet{Barrado2023}.

    \subsection{Bolometric luminosity estimate} \label{res:bollum}
    The broad wavelength range of MIRI/MRS combined with the NIRSpec data provides the possibility to retrieve a bolometric luminosity estimate, as this flux includes the majority of the emergent spectrum of WISE~0855. We integrate the combined spectrum on the original resolution with the empirical and inflated error from 0.8 to 22 µm and add the best fitting ATMO++ model from 22 to 30 µm. We obtain a value of the bolometric luminosity of $\mathrm{log(L/\textup{L}_\odot)} = -7.291 \pm 0.008$.
    
    This value is compatible with the previous estimate including the NIRSpec data from \citet{Luhman2024} of $\mathrm{log(L/\textup{L}_\odot)} = -7.305 \pm 0.020$. The uncertainty on the luminosity from this work is smaller compared to the value stated in \citet{Luhman2024} as it would be expected by adding more wavelength coverage. For the uncertainty estimate we included the 10$^b$ values from the cloudy retrieval.
    
    We can compare the above-stated value to the presented values from the grid model analysis in Table \ref{table:gridmodres_lum}. The median value for the grid models is $\mathrm{log(L/\textup{L}_\odot)} = -7.294 \pm 0.023$, which is consistent to the measured estimate by one sigma. The presented values are integrated by a built-in function in \texttt{species} taking the distance and the fitted radius into account. We integrate the model spectra from 0.45 to 30 µm. We do not calculate this value for the Sonora Elf OWl model, as the model spectra are limited up to 15µm. 

    \begin{figure*}[h]
        \centering
        \includegraphics[width=1\hsize]{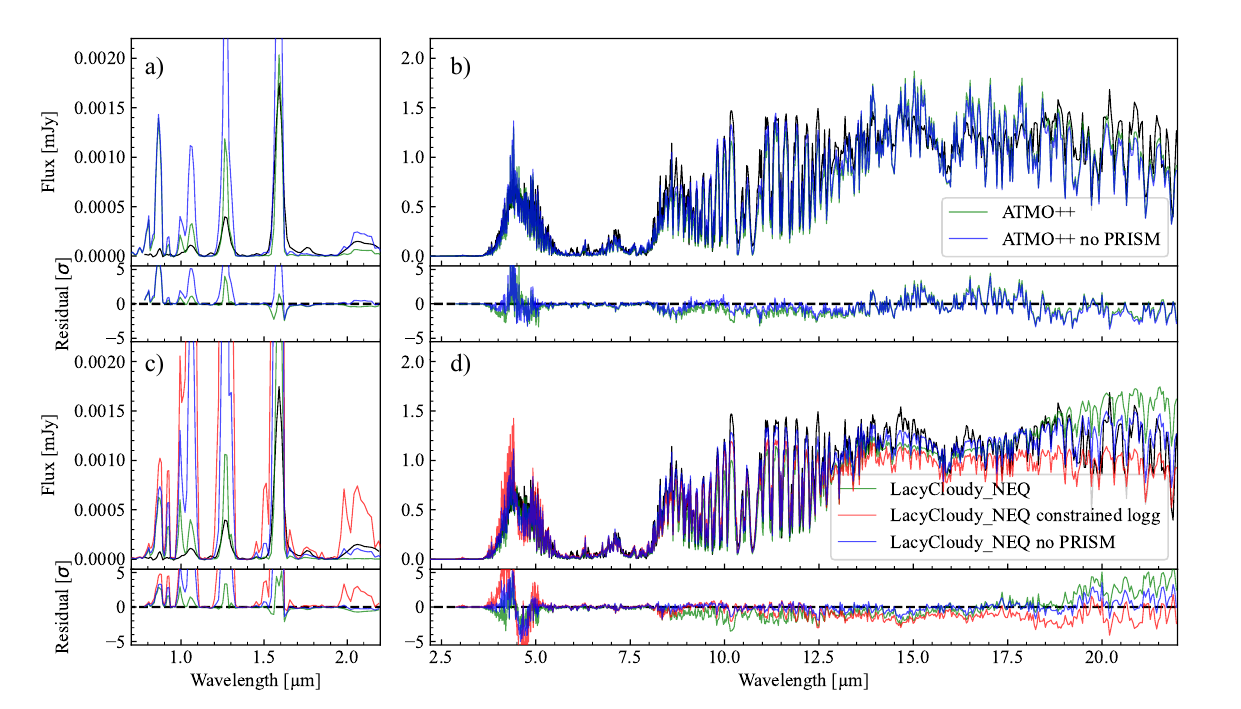}
        \caption{Grid model comparison for ATMO++ in panel a) and b) and Lacy Cloudy non-equilibrium in panel c) and d) if we fit the whole spectrum shown in green versus neglecting the NIRSpec/PRISM and photometry data shown in blue versus constraining the whole spectrum to log(g) between 4.0 and 5.0 in red. In the case for ATMO++ the grid model results do not change if we only constrain the surface gravity and thus we only show the comparison to the neglected NIRSpec/PRISM data. }
    \label{fig:leaveoutprism}
    \end{figure*}

\section{Discussion} \label{discussion} 

    \subsection{Depletion of water in WISE~0855's atmosphere} \label{disc:clouds}
    Due to the low effective temperature of WISE~0855 we expect water to condense out. Using the atmospheric retrievals presented in Section \ref{res:conden} we could detect a lower water abundance in the upper atmosphere of WISE~0855. For both the clear and cloudy case we obtain a Bayes factor which prefers the variable water profile compared to the constant one. Further, when comparing the spectra for constant versus variable water profiles, we see the improvement of the fit at wavelengths larger than 14 µm, shown in Figure~\ref{fig:waterdepl}. This detection shows the power of MIRI/MRS, as both the long wavelengths as well as the medium resolution are needed to resolve this effect.
    
    We were able to connect the lower water abundance in the upper atmosphere to water condensation by comparing the height of abundance decrease with the water condensation and PT profile intersection for the clear retrieval. Water condensation, however, does not directly imply cloud formation and indeed we did not detect any spectral cloud features due to water ice. 
    
    As particles condensate in the atmosphere they become heavier and might sink down to lower atmospheric layers under the large gravity of brown dwarfs. This would lead to a rainout process similar to what is observed for alkali species in L- and T-type objects \citep{Marley2002}. To be able to form a cloud upward mixing is needed to bring warmer air packets to colder regions leading to condensation of species in the packet. Thus, if we see fast settling of the species this might indicate that vertical mixing is weak and stable water ice clouds would not be possible to observe. However, WISE~0855's atmosphere shows disequilibrium chemistry which might be explained by rigours mixing. To understand which effect is the dominant one between upward mixing and settling future variability studies such as the Cycle 1 GO program 2327 may lead to a better understanding of the  mixing timescales and thus the dynamics in WISE0855's atmosphere. Further, the current forward model used in the retrievals might not sufficiently simulate the complex dynamics and interactions with water condensation, cloud formation, rainout processes and many more. 

    Our retrievals constrain the cloud parameters in the cloudy case, setting a cloud deep into the atmosphere. However, this cloud cannot be a water ice cloud due to the high temperatures at the cloud layer base of up to $\sim$500 K. There might be multiple reasons for our findings:

    \begin{itemize}
        \item Adding a cloud to the retrieval adds an opacity source. Thus, this could mimic a missing opacity in our setup that potentially forms a cloud at higher atmospheric pressures. Interestingly, \citet{Morley2018} showed that a low-lying opacity source might better explain the NIR flux for WISE~0855. They proposed the species NH$_4$H$_2$PO$_4$ as a potential candidate. The condensation line of NH$_4$H$_2$PO$_4$ in fact intersects at around $\sim$500 K and 10 bar the modeled PT profile of WISE~0855 as shown in \citet{Morley2018}. By comparison, the deep cloud layer in the retrieval is set at 13 bar, which is close to what they found. NH$_4$H$_2$PO$_4$ may be produced from ammonia-, phosphorus- and water-rich lower atmospheres of gas giants \citep{Visscher2006}. Further, \citet{Beiler2024b} propose that a reason for the missing PH$_3$ in cold brown dwarfs might be a condensation of NH$_4$H$_2$PO$_4$ in the lower atmosphere removing PH$_3$. As we do not detect PH$_3$ in the cloudy retrieval, however constrain PH$_3$ in the clear retrievals, the clouds seem to compensate for adding this species further hinting towards a deep laying opacity source. Thus, we might have revealed a NH$_4$H$_2$PO$_4$ cloud instead of water ice cloud layer in the cloudy retrieval. To test this hypothesis, further investigations on the opacities of this molecule is needed to understand its spectral features. 
        \item Instead of an additional opacity source, we can explain the cloud layer as a compensating effect for a steeper PT structure in the lower atmosphere. This was shown by the fact that the ATMO++ model fitted similarly well compared to the cloudy models by \citet{Lacy2023}.  A change in the PT profile in the lower atmosphere induced by fingering convection processes has been proposed by \citet{Tremblin2015}. For the retrievals the current setup penalizes changes in the PT structure. Thus, setting a deep opacity might be preferred compared to a change in the PT profile in the lower atmosphere. The clear retrievals result in a larger penalty compared to the cloudy one while showing a steeper PT profile. Thus, the penalty might be the reason for setting rather a cloud layer compared to changing the PT structure enforcing this hypothesis. Further, other PT profile setups may be used, such as a slope fitting parameterization similar to the one presented in \citet{Zhang2023}.
        \item Further, the parameterization of the clouds can have an influence on the predicted spectra. \citet{Mang2022} showed that different models for water clouds can lead to significantly different spectra. They compare results from a model including detailed cloud physics with another radiative transfer code based on the parameterization from \citet{Ackerman2001}. They found that they result in incompatible simulated spectra. Thus, the cloud parameterization matters and may strongly vary the resulting spectra. The parameterization used here might be extended in future studies to include more detailed cloud formation processes and structures similarly as presented in \citet{Burningham2021,Vos2023}. In particular the current implementation does not allow for patchy clouds. As discussed in \citet{Morley2014}, water clouds often do not form as a homogeneous cloud deck, but rather heterogeneously with clear and cloudy areas. As we average over such an atmosphere, we would expect to see a linear combination of a clear and cloudy atmosphere. 
    \end{itemize}
    
    Comparing the self-consistent models to the atmospheric retrievals, we find that either clouds as in the models by \citet{Lacy2023} or a PT adaption as in the ATMO++ model is needed to fit the entire spectrum from NIR to MIR. Thus, either one or a combination of the mentioned processes need to be included for fitting WISE~0855's spectrum.

    \subsection{PT structure revisited}
    In our retrieval setup, we do not take any feedback between the PT structure and condensing species into account. The presence of clouds would lead to heat trapping as it is strongly absorbing radiation in the mid infrared, preventing flux from lower in the atmosphere to escape. Consequently, this effect would change the PT structure leading to higher temperatures compared to thermodynamic equilibrium around the cloud base. Also, due to the penalty on the PT profile we potentially miss out on such abrupt changes in the PT structure. Thus, other PT parameterizations accounting for such a change coupled to the cloud layer might be useful to test. \citet{Rowland2023} showed that the parameterization might bias the retrieval results and thus more investigations on the effect of the chosen PT structure on the results is needed. Further, cloudy compared to clear preferences in the retrievals might be driven by PT setup choices as presented in \citet{Whiteford2023}.

    A steeper PT structure in the lower atmosphere was proposed by \citet{Tremblin2015} and could be explained physically by fingering convection. This provides an alternative explanation compared to water ice clouds as we could see in the grid model comparison. Future work on variability measurements might identify patterns only explainable by one of the hypothesis.

    The gravity estimated by the self-consistent models is about 14 times lower compared to the estimate of the retrievals. This might explain the large difference between the PT profile of the \citet{Lacy2023} models and the retrievals, the latter is shifted by about 10 bars towards higher pressures in the pressure range to which the instrument is sensitive. Also the cloudy retrieval with a larger gravity estimate compared to the two clear ones differ in the PT structure. The cloudy retrieval is shifted towards lower atmospheric layers. As \citet{Molliere2015,Fortney2018} showed, offsets in the PT profile can be explained by gravity differences. Potentially, these offsets are degenerate with cloud features and/or metallicity estimates.

    \subsection{Dynamics from chemical disequilibrium} \label{disc:dynam}

    By probing the chemical (dis)equilibrium in brown dwarf atmospheres we can explore the atmospheric dynamics. \citet{Fletcher2019} show how complex it is to model the dynamics in gas giant atmospheres. 
    
    As previously stated by \citet{Miles2020}, CO is present in the atmosphere of WISE~0855, but would not be abundant in the upper atmosphere in chemical equilibrium as shown in Figure \ref{fig:abund_diseq}. Thus, this hints towards mixing processes in the atmosphere.
    The eddy diffusion coefficient quantifies the strength diffusive mixing. Here, our retrieval finds a value of log(K$_{zz}$) = 1.3 $\pm$ 0.2 cm$^2$/s. This is significantly different from what \citet{Miles2020} found by comparing the M band spectrum to grid models: log(K$_{zz}$) = 8.5 cm$^2$/s corresponding to a much larger diffusion compared to the retrieval estimate. \citet{Leggett2021} confirm the estimate by \citet{Miles2020} by obtaining a value log(K$_{zz}$) = 8.7 cm$^2$/s. From Sonora Elf Owl we get a value of log(K$_{zz}$) = 2.25 $\pm$ 0.04 cm$^2$/s, which is larger compared to the value from the retrieval, however smaller than the estimates by \citep{Miles2020} and \citep{Leggett2021}. Our retrieval estimate however is based on a value derived only from the particle size of the retrieved water cloud and as we do not detect the water ice cloud, this value should be taken with caution.
    More variability measurements similar to the work presented by \citet{Esplin2016} may give better constraints on the dynamic processes. Also, by modelling well known objects, such as Jupiter, we might obtain narrower boundary values on priors helping us to better constrain abundances and bulk parameters and to rule out certain scenarios.
   
    From chemical equilibrium we would expect more $\mathrm{NH_3}$ compared to $\mathrm{N_2}$ in the photosphere \citep{Saumon2006}. However if we are in a disequilibrium we would observe a decrease in the $\mathrm{NH_3}$ abundance \citep{Leggett2021,Morley2018}. We detect lower values of NH$_3$ compared to chemical equilibrium estimates. This is a further indication for disequilibrium chemistry. A smaller value of ammonia might also indicate that nitrogen containing species are condensing out in lower layers of the atmospheres. A candidate molecule could be the previously mentioned NH$_4$H$_2$PO$_4$. Depletion of ammonia was identified by \citep{Leggett2015}. 
    For methane we find values as they were expected from chemical equilibrium calculations. However, this might not be constant throughout the atmosphere and \citet{Morley2018} see indications for a methane depletion in WISE~0855. 
    
    In the presented work, except for water, we use constant abundances of trace molecules with height. This is a fair assumption when probing a small wavelength range, as we are then probing only a small part of the atmosphere. However, in our case it is a very strong assumption as we probe the atmosphere from about 100 to 0.3 bar and we can expect the abundance to vary in this broad height range. Analysis shows how changes in the abundances with height can significantly affect the spectra  \citep{Rowland2023}. In future retrievals, variable profiles for molecules other than water may be included, especially when analyzing broad wavelength ranges.

    \subsection{Evolution of the coldest brown dwarf} 
    \citet{Leggett2021} show their evolutionary models and give estimates for WISE~0855. We can compare our results with their Figure 11. A temperature of about 300 K and a gravity of log(g) = 4.7 and 4.9 for the clear and cloudy case respectively would lead to ages larger than 10 Gyr, masses of larger than 20 M$_J$ and radii smaller than 1 R$_J$. In contrast, the estimates from the self-consistent models with a log(g) of 3.5 and a effective temperature of 250 K would result in masses of about 1 M$_J$, ages of about 0.3 Gyr and radii of about 1.1 R$_J$. This is compatible with what we find for the Lacy models. 
    When comparing our values to the evolutionary model presented in \citet{Marley2021}, we get similar values as presented in \citet{Leggett2021}. For the effective temperature of about 250 K and log(g) of 3.5, we obtain a bolometric luminosity of about -7.4, a mass of 1.5 M$_J$, an age of 0.4 Gyr and a radius of 1.2 R$_J$. The bolometric luminosity is comparable  however slightly smaller compared to our estimate.
    
    Measuring the ratio between $\mathrm{^{14}NH_3}$/$\mathrm{^{15}NH_3}$ has been identified as a new formation tracer \citep{Barrado2023}. As discussed in Section 4.7 we report a value for the $\mathrm{^{14}NH_3/^{15}NH_3}$ ratio of 332 $^{+63}_{-43}$ for the clear retrieval, being closer to the ISM value than the value for the Solar System or WISE~1828 of 670$^{+390}_{-211}$. This means the atmosphere of WISE~0855 seems to be more enriched in $\mathrm{^{15}NH_3}$ compared to WISE~1828. Thus, WISE~0855 might have formed at different initial conditions of the molecular cloud compared to WISE~1828. Potentially it could have been a younger cloud, as the ratio resembles more today's enriched ISM value compared to an enrichment as in the several Gyr old Solar System. More measurements of isotopologues in brown dwarfs are needed to identify trends and to be able to connect to formation conditions and evolutionary models.

    \subsection{Challenges in consistently fitting the data}
    Generally, it is difficult for the self-consistent grid models to fit the entire spectrum across the broad wavelength range. To illustrate this, we show in Figure \ref{fig:leaveoutprism} the ATMO++ (upper plots) and the cloudy Lacy non-equilibrium fits (lower plots). We present the fits over the entire wavelength range in green and the fit over only NIRSpec/G395M and MIRI/MRS in blue. In both cases, but more prominently in the Lacy fit, the model agrees better with the mid-infrared data if we neglect NIRSpec/PRISM. However, we then obtain an overshoot in the NIR wavelength range. Thus, this indicates that besides a missing opacity source or change in the PT structure even another yet unaccounted effect is needed to explain the full spectrum. 
    The red line shows the cloudy Lacy model with the constrained surface gravity to log(g) = 4.0 to 5.0. In comparison with the unconstrained fit this shows the impact of a higher gravity on the spectrum. However, constraining the gravity does not improve the fit nor does it find another local minimum in which the gravity might be larger as the constrained fit also reaches the lower end of the prior. 
    
    How well we fit the data for the grid models per data set across the wavelength range is presented in Figure \ref{fig:chisquared} in the appendix. The cloudy non-equilibrium models by \citet{Lacy2023} show the smallest $\chi^2$ values for each of the data sets. Comparably well fits the ATMO++ model with similar values as the lowest cloudy grid fit especially in the mid infrared. Sonora Elf Owl fits the NIRSpec/PRISM data below 2.2 µm well, but shows larger $\chi^2$ for larger wavelengths. Both clear models by Lacy fit significantly worse above 2.2 µm compared to the other models. Note that the overall offsets per data set are influenced by the retrieved $10^b$ factor added to the error, which varies for each data set.
    
    Changing the error compared to the pipeline error requires careful consideration as the resulting fit and statistical interpretation depend on this. As discussed in detail in Section \ref{data}, we inflated the error on the data. The solution presented by \citep{Line2015} by adding an additional offset to the error estimate which is a free parameter of the retrieval works well and has been used in multiple analyses already \citep[e.g.][]{Barrado2023,Kothari2024,Rowland2023}. However, we obtain generally very small width in the posterior distributions leading to small uncertainties on the reported values. The resulting width of the posterior distribution might be influenced by the error inflation performed on the fitted data set. Thus, to better understand where the small errors arise from, further investigations on the impact of the error inflation on the posteriors are needed.

    As shown in \citep{Rowland2023} the comparison of free atmospheric retrievals and self-consistent grid models helps to set the results of each method into perspective. However, both methods differ in their assumptions and this needs to be taken into account when comparing them. For example we assume constant abundances across the studied pressure range, where self-consistent models vary the abundances. Self-consistent grid models fit only a few parameters compared to retrievals fitting for several tens of parameters. Thus, the fits of retrievals are often better, sometimes even tending to over-fit the data.

\section{Summary and Outlook} \label{conclusion}
In this paper, we present the mid infrared spectrum of WISE~0855 from 5 to 22 µm observed by JWST MIRI/MRS and a combined analysis with the MIRI/MRS, NIRspec/G395M and NIRSpec/PRISM data \citep{Luhman2024}. We discuss how the mid-infrared adds to existing knowledge of WISE~0855's atmospheric composition and structure using atmospheric retrievals and self-consistent grid models.

\begin{itemize}
    \item Atmospheric retrievals reveal a depletion of water in the upper atmosphere of WISE~0855 and we could link it to the condensation of water.
    \item We could not detect clear spectral signatures of water ice particles in the atmosphere of WISE~0855 using either atmospheric retrievals or self-consistent grid models. The cloudy retrieval constrains the cloud parameter, however the retrieved cloud layer is located too deep in the atmosphere to form stable water ice clouds. This cloud however potentially mimics a phosphorous cloud formed by the proposed species $\rm NH_4H_2PO_4$.
    \item Either a cloud layer or an adaption in the PT profile needs to be included in the self-consistent grid models to explain WISE~0855's spectrum.
    \item The cloudy retrieval results in slightly larger equilibrium chemistry values for CH$_4$, significantly larger values for CO, slightly smaller values for NH$_3$ and unconstrained values of PH$_3$ and H$_2$S. Similar values are retrieved in the clear retrievals for all species except for PH$_3$ and H$_2$S. 
    \item We detect $\mathrm{^{15}NH_3}$ in WISE~0855 and measure its abundance using atmospheric retrievals. The isotopologue ratio is determined as 332$^{+63}_{-43}$ for the clear and variable profile retrieval.
    \item We calculate a bolometric luminosity of $\mathrm{log(L/\textup{L}_\odot)} = -7.291 \pm 0.008$ by integrating the spectrum and best-fit ATMO++ model from 0.8 to 30 µm.
\end{itemize}

The MIRI/MRS observation presented in this study, enables new possibilities in the atmospheric characterization of WISE~0855 by covering the mid infrared wavelengths in medium resolution, as shown by the detection of water depletion. The implementation of liquid water in the retrievals could lead to a better understanding of the condensation processes. Potentially missing deep opacity sources may be identified using future retrievals and modeling approaches. Future observations could reveal WISE~0855's atmospheric variability, potentially identifying patchy clouds or better constraining the atmospheric dynamics and pressure temperature structure. Y dwarfs are fascinating analogues to gas giants both in and beyond our Solar System. MIRI allows their detailed characterization leading to a better understanding of the very low temperature regime of brown dwarf atmospheres. 

\begin{acknowledgements}
    P.-O.L acknowledges funding support by CNES.
    BV, OA, and PR thank the European Space Agency (ESA) and the Belgian Federal Science Policy Office (BELSPO) for their support in the framework of the PRODEX Programme.
    OA is a Senior Research Associate of the Fonds de la Recherche Scientifique – FNRS.
    EvD acknowledges support from A-ERC grant 101019751 MOLDISK.
    DB is supported by Spanish MCIN/AEI/10.13039/501100011033 grant PID2019-107061GB-C61 and and No. MDM-2017-0737.
    LD acknowledges funding from the KU Leuven Interdisciplinary Grant (IDN/19/028), the European Union H2020-MSCA-ITN-2019 under Grant no. 860470 (CHAMELEON) and the FWO research grant G086217N.
    MPIA acknowledges support from the Federal Ministry of Economy (BMWi) through the German Space Agency (DLR).
    PP thanks the Swiss National Science Foundation (SNSF) for financial support under grant number 200020\_200399.
    PT acknowledges support by the European Research Council under Grant Agreement ATMO 757858.
    JPP acknowledges financial support from the UK Science and Technology Facilities Council, and the UK Space Agency.
    For the purpose of open access, the author has applied a Creative Commons Attribution (CC BY) licence to the Author Accepted Manuscript version arising from this submission.
    The Cosmic Dawn Center (DAWN) is funded by the Danish National Research Foundation under grant No. 140. 
    TPR acknowledges support from the ERC through grant no.\ 743029 EASY.
    MV is a FRIA grantee of the Fonds de la Recherche Scientifique – FNRS.
    GO acknowledges support from SNSA.
\end{acknowledgements}

%
%

\bibliographystyle{aa} 
\bibliography{export} 

\begin{appendix}
\section{ }

    \begin{figure*}[h] 
   \centering
   \includegraphics[width=1\textwidth]{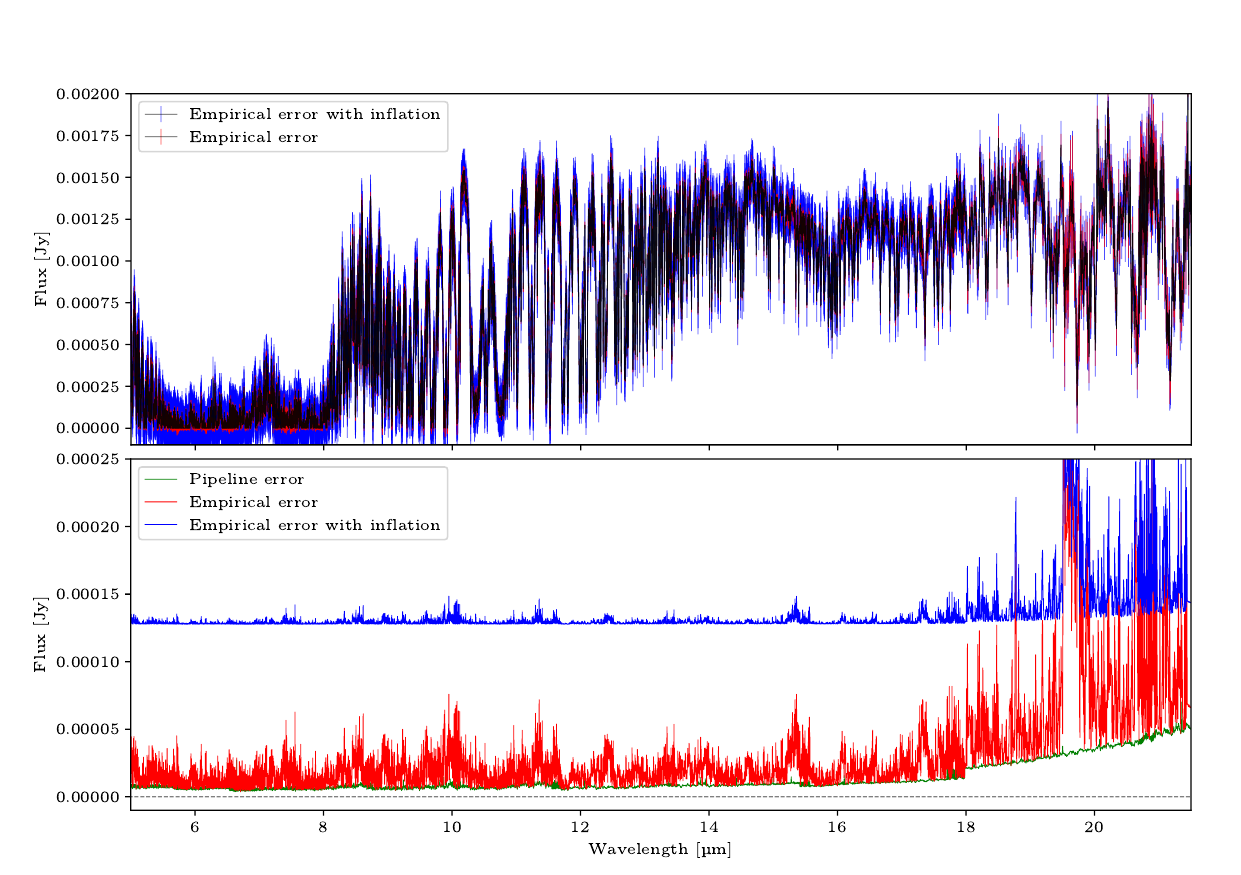}
   \caption{The upper plot shows the data with the uncertainty either estimated from the empirical error shown in red error bars and with the error inflation factor in blue error bars. The lower plot shows the error estimated from the pipeline, the empirical error and the final error with the included error inflation factor plotted along the wavelength. The inflation factor used here is $b = -7.78$.}
    \label{fig:errors}
    \end{figure*}

    \begin{figure*}[h]
   \centering
   \includegraphics[width=0.95\textwidth]{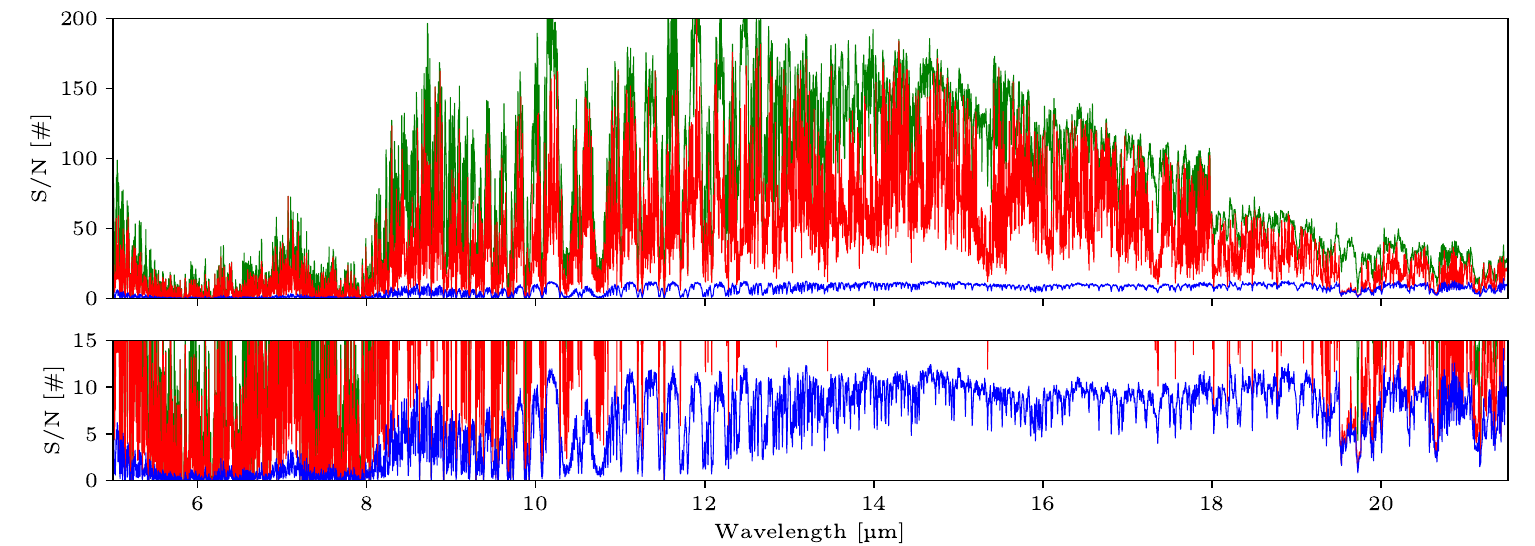}
   \caption{The signal to noise ratio (SNR) for the error of the pipeline in green, the empirical error in red and the empirical error with the inflation factor in blue. The lower plot shows an inset of the upper plot scaled in the y-axis to better visualize the SNR of the empirical error with the inflation factor. The latter data are the data used in the analysis.}
    \label{fig:snr}
    \end{figure*}

    \begin{figure*}[h] 
   \centering
   \includegraphics[width=1.05\textwidth]{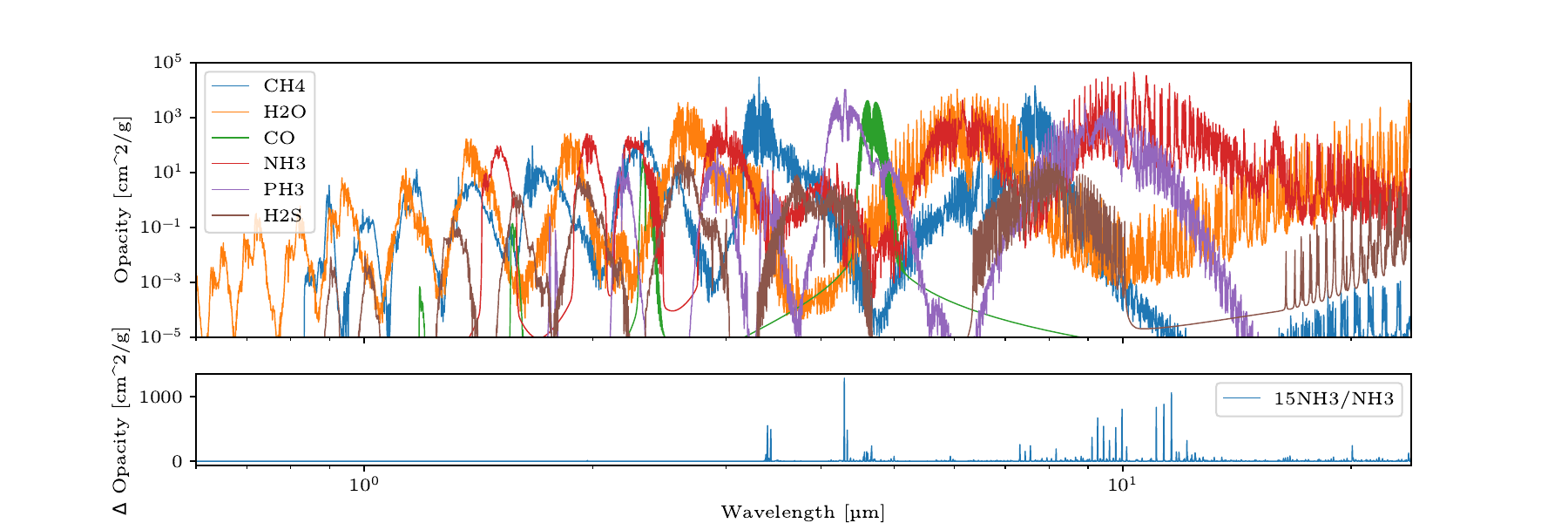}
   \caption{The upper plot shows the log opacities of methane, water, carbon monoxide, ammonia and phosphine for the retrieval analysis. The lower plot shows the ratio between the opacity of $\mathrm{^{15}NH_3}$ and NH$_3$ as a function of wavelength showing where we would expect changes in the opacity due to the isoptopologue. The opacities were taken at the temperature of T$_{\rm eff}$ = 276 K and a pressure of 1.02 bar according to the effective temperature estimate of the retrieval and the corresponding PT profile. }
   \label{fig:opacities}
    \end{figure*}

    \begin{figure*}[h]
   \centering
   \includegraphics[width=0.95\textwidth]{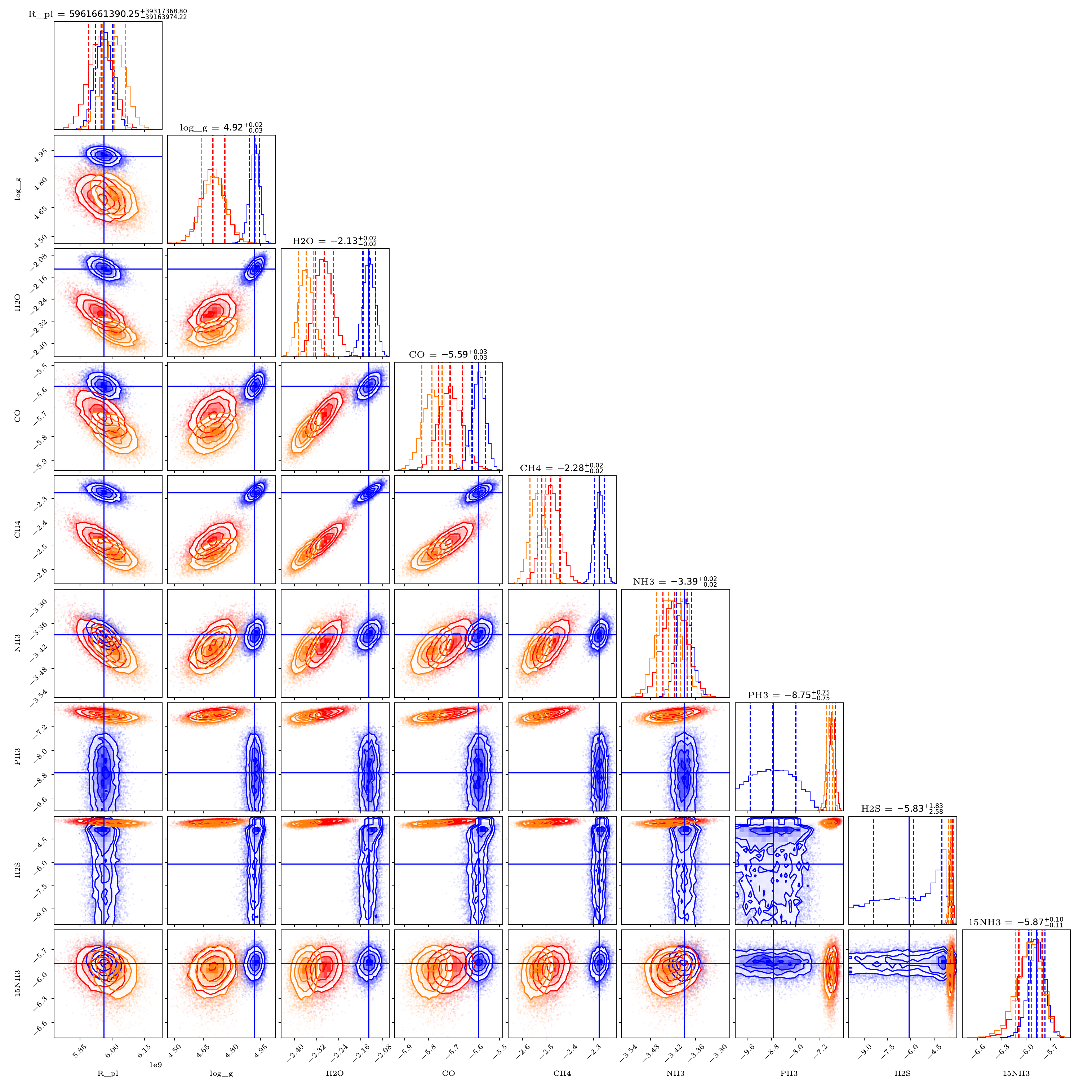}
   \caption{Posterior distributions of the retrieval outputs for the Radius (R\_pl) in cm, the surface gravity (log(g)) in cm/s$^2$ and the logarithmic mass fractions of the following molecules: $\mathrm{H_2O}$, $\mathrm{CO}$, $\mathrm{CH_4}$, $\mathrm{NH_3}$, $\mathrm{PH_3}$, $\mathrm{H_2S}$ and $\mathrm{^{15}NH_3}$. In blue we show the results from the cloudy, variable water profile retrieval, in red the distributions from the clear and variable water profile and in orange the values of the clear and constant water profile retrieval. The black lines correspond to the mean values for the cloudy retrieval.}
    \label{fig:corner_abund}
    \end{figure*}

    \begin{figure*}[h]
   \centering
   \includegraphics[width=0.95\textwidth]{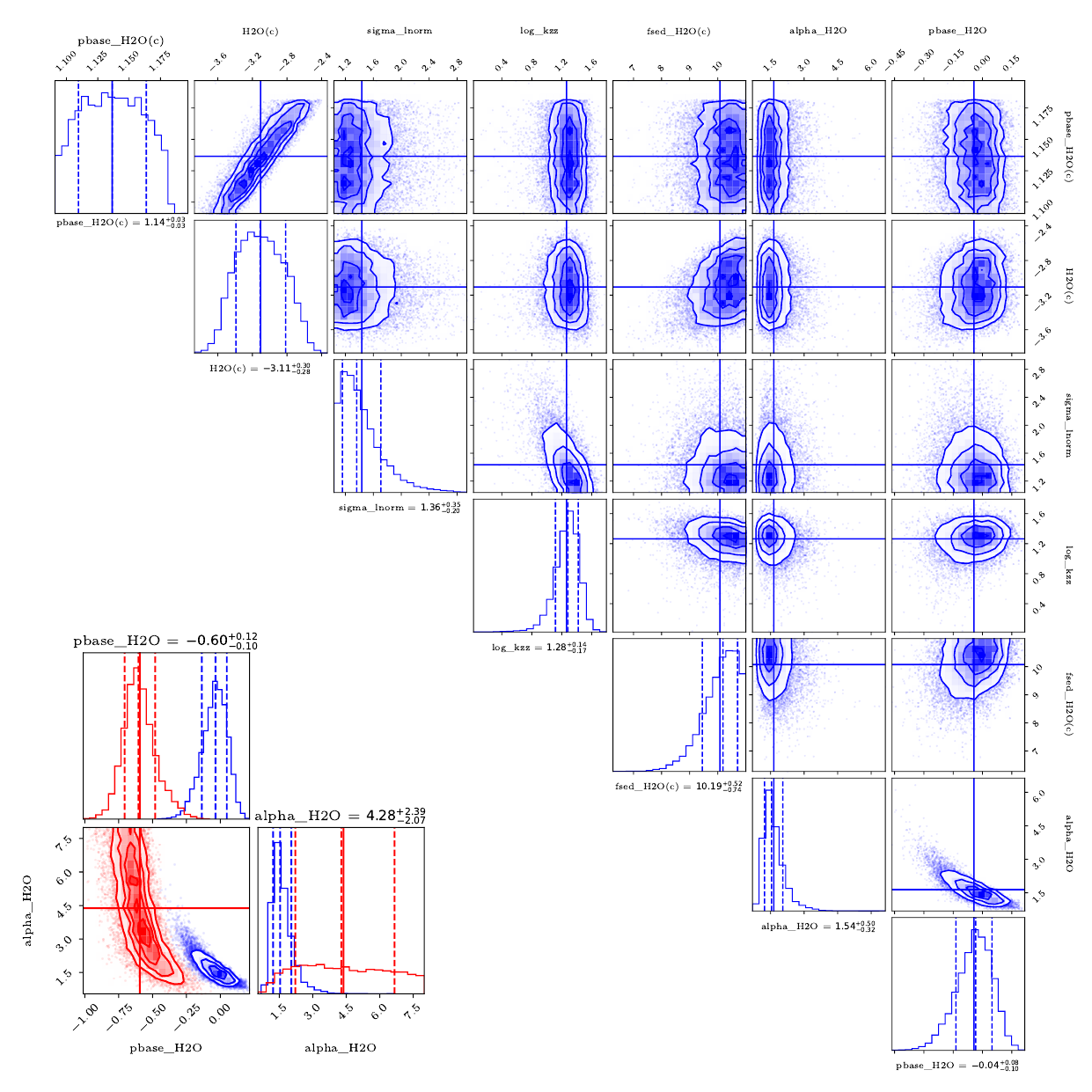}
   \caption{The posterior distributions for the cloud parameter. On the upper reversed plot, we show the following variables for the cloudy with variable water profile retrieval: The logarithmic pressure for the cloud base (log\_Pbase\_H2O(c)), the logarithmic cloud particle mass fraction (log\_X\_cb\_H2O(c)), the width of the cloud particle distribution (sigma\_lnorm), the logarithmic eddy diffusion coefficient (log\_kzz), the sedimentation parameter and cloud thickness indicator (f\_sed), the amount by which the liquid water in the atmosphere is reduced by the variable water profile (alpha\_H2O), the pressure value where to reduce the amount of liquid water (pbase\_H2O). The lower plot compares the two values for the variable water profile (pressure and amount of change in abundance) for the two retrievals with variable water profiles: In blue we show the distribution for the cloudy and in red for the clear retrieval. The values noted on top of each distribution correspond to the median and the 16th and 84th percentile correspondingly. The solid lines correspond to the mean values of the cloudy and the red lines to the ones of the clear retrieval with variable water profile.}
    \label{fig:corner_cloud}
    \end{figure*}

\begin{table*}
\caption{Retrieval results of the three presented retrievals. We compare the posterior distribution results to the posteriors, the best fit values and show the uniformly sampled priors used for the setup. The posterior values correspond to the median values on the posterior distribution and the uncertainty estimates on the difference between the median and the 16th and 84th percentile correspondingly. For molecules we report logarithmic mass fractions.
}             
\label{table:posteriors}      
\centering          
\begin{tabular}{c c c c c l l l }     
\hline\hline       
Variable & Prior & \multicolumn{2}{c}{Retrieval clear const. H$_2$O} & \multicolumn{2}{c}{Retrieval clear var. H$_2$O} & \multicolumn{2}{c}{Retrieval cloudy var. H$_2$O}\\ 
\hline                  
   & & Posterior & Best Fit & Posterior & Best Fit & Posterior & Best Fit \\  \hline \hline  
   Radius [R$_J$] & $\mathcal{U}$(0.489, 2.934) & 0.832$^{+0.008}_{-0.008}$  & 0.837 & 0.832$^{+0.008}_{-0.008}$    & 0.833 & 0.833$^{+0.005}_{-0.005}$ & 0.837\\
   log(g) [cm/s$^2$] & $\mathcal{U}$(2.0,6.0) & 4.70$^{+0.06}_{-0.06}$  & 4.69 & 4.70$^{+0.06}_{-0.06}$& 4.70 & 4.92$^{+0.02}_{-0.03}$ & 4.94\\
   M [M$_{\rm J}$] & -  & 14.28$^{+2.05}_{-2.13}$  &  13.85 & 14.14$^{+1.77}_{-1.87}$  &   14.03 & 23.29$^{+1.29}_{-1.31}$  &  24.59 \\
   T$_{eff}$ [K] & -  & 296.8$^{+1.5}_{-1.3}$  &  299.2 & 298.3$^{+1.6}_{-1.5}$  &   299.9  & 297.7$^{+1.1}_{-1.0}$  &  299.0 \\
   log(H$_2$O) & $\mathcal{U}$(-9.999999, -0.000001)  & -2.36$^{+0.03}_{-0.03}$     & -2.37 & -2.29$^{+0.03}_{-0.03}$ & -2.30 & -2.13$^{+0.02}_{-0.02}$ & -2.12 \\
   log(CO) & $\mathcal{U}$(-9.999999, -0.000001)      & -5.78$^{+0.04}_{-0.04}$    & -5.76 & -5.71$^{+0.05}_{-0.05}$ & -5.71 & -5.59$^{+0.03}_{-0.03}$ & -5.56\\
   log(CH$_4$) & $\mathcal{U}$(-9.999999, -0.000001)  & -2.54$^{+0.03}_{-0.03}$    & -2.53 & -2.48$^{+0.04}_{-0.04}$& -2.48 & -2.28$^{+0.02}_{-0.02}$ & -2.27\\
   log(NH$_3$) & $\mathcal{U}$(-9.999999, -0.000001)  & -3.43$^{+0.03}_{-0.03}$  & -3.42 & -3.41$^{+0.03}_{-0.03}$& -3.41 & -3.39$^{+0.02}_{-0.02}$ & -3.40\\
   log(PH$_3$) & $\mathcal{U}$(-9.999999, -0.000001)  & -6.88$^{+0.09}_{-0.10}$    & -6.89 & -6.79$^{+0.09}_{-0.10}$& -6.78 &  -8.75$^{+0.75}_{-0.75}$ & -8.56 \\
   log(H$_2$S) & $\mathcal{U}$(-9.999999, -0.000001)  & -3.46$^{+0.09}_{-0.10}$    & -3.53 & -3.37$^{+0.09}_{-0.10}$& -3.42 & -5.83$^{+1.83}_{-2.58}$ & -3.79 \\
   log($\mathrm{^{15}NH_3}$) & $\mathcal{U}$(-9.999999, -0.000001)  & -5.97$^{+0.14}_{-0.18}$  & -6.08 & -5.94$^{+0.15}_{-0.16}$ & -5.90 & -5.87$^{+0.10}_{-0.11}$ & -5.82\\
   $\gamma$ & $\mathcal{U}$(0.000000, 9999953948)     & 2.9e-7$^{+0.5}_{-0.4}$    & 2.5e-7 & 3.4e-7$^{+0.6}_{-0.5}$  & 3.8e-7 & 6.6e-7$^{+1.0}_{-0.9}$  & 7.0e-7\\
   log(pbase$_{H_2O}$) & $\mathcal{U}$(-5.999999, 3.999999)  & -    & - & -0.60$^{+0.11}_{-0.10}$  & -0.71 & -0.04$^{+0.08}_{-0.10}$ & -0.04\\
   alpha$_{H_2O}$ & $\mathcal{U}$(-0.999999,7.999999)             & -    & - & 4.28$^{+2.29}_{-2.07}$ & 6.77 & 1.54$^{+0.50}_{-0.32}$ & 1.27 \\
   f$_{\rm sed}$ & $\mathcal{U}$(1.000001, 10.999999)     & -    & - & -& - & 10.2$^{+0.5}_{-0.7}$ & 10.2 \\
   log(K$_{zz}$) [cm$^2$/s] & $\mathcal{U}$(0.000001, 9.999999)  & -    & - & -& - & 1.3$^{+0.1}_{-0.2}$ & 1.4 \\
   log($\sigma_{\rm lnorm}$) & $\mathcal{U}$(1.01,2.96)   & -    & - & -& - & 1.39$^{+0.37}_{-0.22}$ & 1.27 \\
   b$_{\rm MIRI}$ & $\mathcal{U}$(-13.400955, -5.638476)  & -7.71$^{+0.02}_{-0.02}$    & -7.72 & -7.73$^{+0.02}_{-0.02}$ & -7.73 & -7.77$^{+0.01}_{-0.01}$ & -7.76\\
   b$_{\rm G395M}$ & $\mathcal{U}$(-18.697953, -7.392405) & -8.77$^{+0.02}_{-0.02}$    & -8.76 & -8.76$^{+0.03}_{-0.02}$ & -8.76 & -8.79$^{+0.02}_{-0.02}$ & -8.78\\
   b$_{\rm PRISM1}$ & $\mathcal{U}$(-19.618027, -8.789751)& -13.41$^{+0.07}_{-0.07}$    & -13.48 & -13.48$^{+0.08}_{-0.08}$ & -13.41 &  -14.08$^{+0.06}_{-0.06}$ & -14.11\\
   b$_{\rm PRISM2}$ & $\mathcal{U}$(-20.736958, -12.267634)& -9.54$^{+0.08}_{-0.08}$    & -9.57 & -9.50$^{+0.08}_{-0.08}$ & -9.51 & -9.89$^{+0.08}_{-0.08}$ & -9.88\\
   log(H$_2$O(c)) & $\mathcal{U}$(-9.999999, -0.000001)& -    & - & -& - & -3.11$^{+0.30}_{-0.28}$ & -3.33\\
   log($\mathrm{pbase_{H_2O(c)}}$) & $\mathcal{U}$(-5.0, 2.0)  & -    & - & -& - & 1.14$^{+0.03}_{-0.03}$ & 1.11 \\
   
\hline                  
\end{tabular}
\end{table*}

 \begin{table*}
    \caption{The nested sampling global logarithmic evidence for the presented atmospheric retrievals. }             
    \label{table:goodnessfit_ret}      
    \centering                          
    \begin{tabular}{l l l l}        
    \hline\hline                 
    Model & ln(Z)  \\    
    \hline                        
       Ret. clear, const. H$_2$O    & 20460.75 $\pm$ 0.29 \\ 
       Ret. clear, var. H$_2$O      & 20468.43 $\pm$ 0.30 \\
       Ret. cloudy, var. H$_2$O     & 20601.93 $\pm$ 0.33 \\
       Ret. cloudy, const. H$_2$O.  & 20567.90 $\pm$ 0.31 \\
       Ret. clear, const. H$_2$O, no $\mathrm{^{15}NH_3}$ & 20450.64 $\pm$ 0.29 \\
    \hline                                   
    \end{tabular}
    \end{table*}

    \begin{figure*}[h] 
    \centering
    \includegraphics[width=1\textwidth]{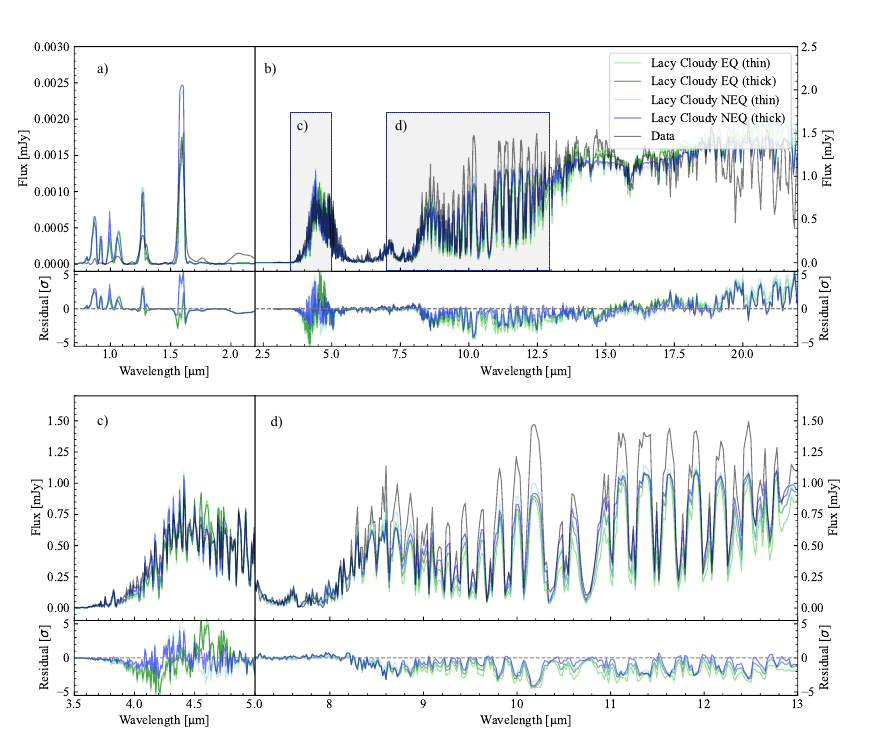}
    \caption{Comparing the cloudy models both with and without equilibrium chemistry and two different thicknesses of water clouds by \citet{Lacy2023}. The best fit models in the cloudy case using the the equilibrium thin and thick and the disequilibrium self-consistent thin and thick grid models by \citet{Lacy2023}. The corresponding reduced $\chi^2$ values are 2.48, 1.97, 3.48 and 2.75 for the equilibrium thin, thick and disequilibrium thin and thick cloudy models. As for the other models, we re-binned the models and data to a resolution R = 500 for NIRSpec/G395M and MIRI/MRS and NIRSpec/PRISM is kept at R = 100.}%
    \label{fig:selfcons_chem_cloudy}
    \end{figure*}

    \begin{figure*}[h] 
    \centering
    \includegraphics[width=1\textwidth]{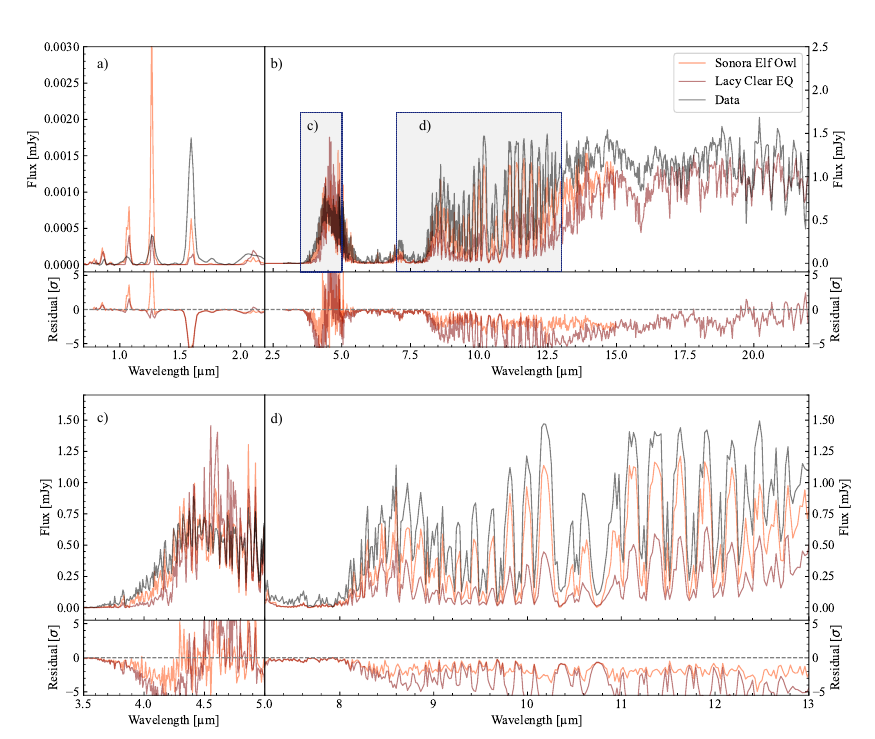}
    \caption{Comparing the best-fit equilibrium chemistry clear model by \citet{Lacy2023} with the clear disequilibrium model Sonora Elf Owl by \citet{Mukherjee2024}. The corresponding reduced $\chi^2$ is 12.72 for the Lacy model and 6.77 for Sonora Elf Owl. Again, for better visibility the presented models and data are re-binned to a resolution R = 500 for the NIRSpec/G395M and MIRI/MRS and NIRSpec/PRISM is kept at  R = 100.}
    \label{fig:selfcons_chem_clear}
    \end{figure*}

     \begin{figure*}[h]
   \centering
   \includegraphics[width=0.95\textwidth]{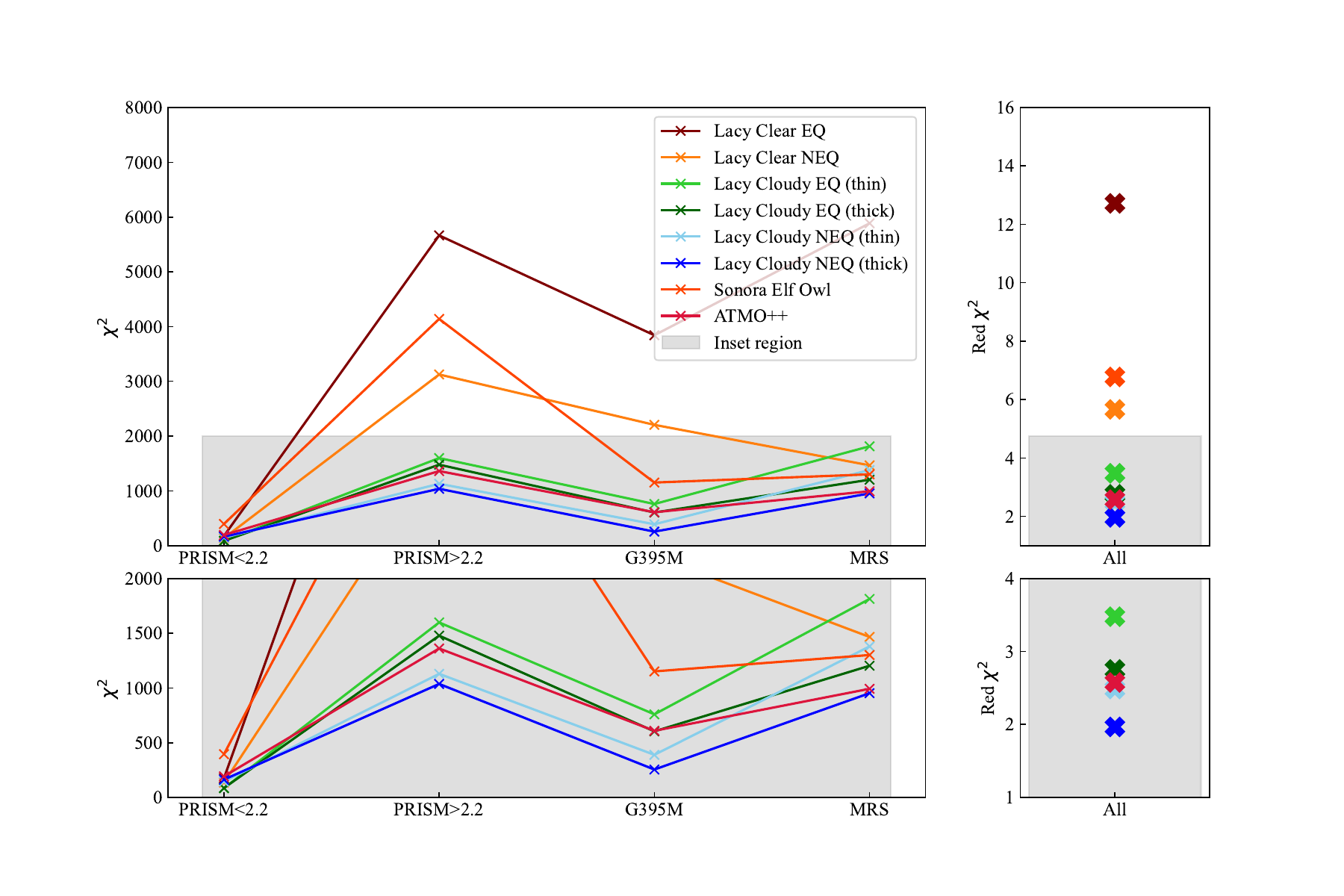}
   \caption{The left plots compare the $\chi^2$ values for the different self-consistent models for each part of the spectrum. The right plots show the overall reduced $\chi^2$ for each model. The upper plots show all models and the lower plot show an inset region marked in grey. The Lacy Cloudy NEQ (thick) model reaches the lowest $\chi^2$ values over all. The ATMO++ model fits similarly well especially in the mid infrared. We use the values of the $10^b$ from the clear and variable water profile retrieval for the $\chi^2$ calculation.}
    \label{fig:chisquared}
    \end{figure*}

    \begin{figure*}[h] 
    \centering
    \includegraphics[width=1\textwidth]{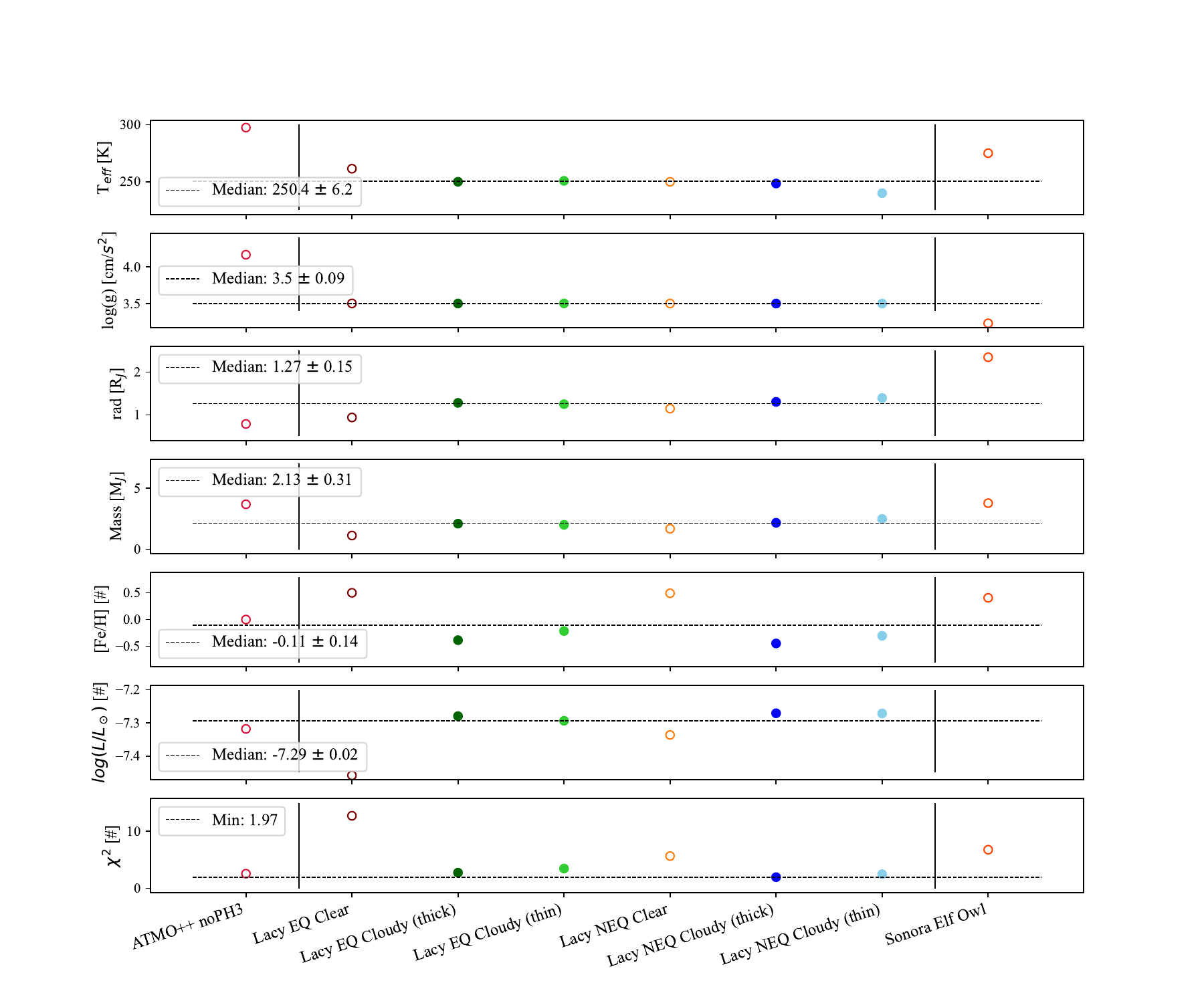}
   \caption{Comparing the model value outputs of the self-consistent grid model ATMO++ \citep{Leggett2021,Meisner2023}, \citet{Lacy2023} and Sonora Elf Owl \citet{Mukherjee2024} from top to bottom: The effective temperature $T_{eff}$, the logarithmic surface gravity log(g), the radius of the gas giant, the calculated mass based on the radius and the surface gravity, the metallicity [Fe/H] with 0 solar values and +/- 0.5 super and sub-solar values correspondingly, the logarithmic luminosity compared to solar and the $\chi^2$ of the fit. Filled circles correspond to cloudy models and the colors correspond to the same as shown in Figures \ref{fig:cloud_self},\ref{fig:selfcons_chem_cloudy} and \ref{fig:selfcons_chem_clear}. Errors of the estimated values are too small to be depicted. The dotted lines correspond to the median value of the eight shown model spectra and in the legend the median and the standard error are presented. }%
   \label{fig:selfcons_comp}
    \end{figure*}
    
    \begin{table*}[h]
    \caption{Summary of the fitted parameters of the self-consistent grid models: ATMO++, Lacy \& Burrow and Sonora Elf Owl. }  
    \label{table:gridmodres}      
    \centering          
    \begin{tabular}{l l l l l l l l l  }     
    \hline\hline       
    Model & $\mathrm{T_{eff}}$ [K] & log(g) [cm/s$^2$] & $\mathrm{R_{pl}}$ [$\times$ R$_J$] & Mass [$\times$ M$_J$] &[M/H] [\#]  \\ 
    \hline                    
       ATMO ++ no PH$_3$            & 297.4 $\pm$ 0.7       & 4.167 $\pm$ 0.008    & 0.789 $\pm$ 0.006 & 3.6809 $\pm$ 0.0003 & 0                  \\
       Lacy eq. clear               & 261.4 $\pm$ 0.6       & 3.501 $\pm$ 0.001    & 0.940 $\pm$ 0.009 & 1.1300 $\pm$ 0.0004 & 0.499 $\pm$ <0.001  \\
       Lacy eq. cloudy (thick)      & 250.0 $\pm$ < 0.1     & 3.500 $\pm$ < 0.001  & 1.281 $\pm$ 0.009 & 2.0966 $\pm$ 0.0007 & -0.386 $\pm$ 0.002  \\ 
       Lacy eq. cloudy (thin)       & 250.8 $\pm$ 0.2       & 3.501 $\pm$ 0.001    & 1.251 $\pm$ 0.008 & 2.0037 $\pm$ 0.0006 & -0.215 $\pm$ 0.002  \\ 
       Lacy non-eq. clear           & 250.0 $\pm$ 0.1       & 3.500 $\pm$ < 0.001  & 1.147 $\pm$ 0.008 & 1.6803 $\pm$ 0.0005 & 0.491 $\pm$ 0.003  \\
       Lacy non-eq. cloudy (thick)  & 248.4 $\pm$ 0.5       & 3.500 $\pm$ < 0.001  & 1.304 $\pm$ 0.011 & 2.1693 $\pm$ 0.0007 & -0.447 $\pm$ 0.004  \\
       Lacy non-eq. cloudy (thin)   & 240.0 $\pm$ 0.5       & 3.500 $\pm$ < 0.001  & 1.395 $\pm$ 0.011 & 2.4826 $\pm$ 0.0008 & -0.305 $\pm$ 0.004 \\
       Sonora Elf Owl               & 275.0 $\pm$ < 0.1     & 3.230 $\pm$ < 0.001  & 0.975 $\pm$ 0.003  & 0.3252 $\pm$ 0.0004   & 0.406 $\pm$ 0.005    \\ 
      \hline           
      \textit{Ignore NIRSpec/PRISM and phot.} \\
      \hline

       ATMO ++ no PH$_3$            & 309.0 $\pm$ 0.6       & 4.094 $\pm$ 0.008   & 0.742 $\pm$ 0.006 & 2.7582 $\pm$ 0.0002 & 0                  \\
       Lacy eq. clear               & 256.5 $\pm$ 0.6       & 3.501 $\pm$ < 0.001 & 1.002 $\pm$ 0.009 & 1.2839 $\pm$ 0.0004 & 0.499 $\pm$ < 0.001  \\
       Lacy eq. cloudy (thick)      & 284.1 $\pm$ 5.3       & 4.000 $\pm$ < 0.001 & 1.376 $\pm$ 0.260 & 7.6418 $\pm$ 0.0011 & -0.127 $\pm$ 0.002  \\ 
       Lacy eq. cloudy (thin)       & 293.6 $\pm$ 1.1       & 3.745 $\pm$ 0.008   & 0.899 $\pm$ 0.013 & 1.8144 $\pm$ 0.0003 & -0.057 $\pm$ 0.003 \\ 
       Lacy non-eq. clear           & 278.4 $\pm$ 0.2       & 3.500 $\pm$ < 0.001 & 0.933 $\pm$ 0.006 & 1.1127 $\pm$ 0.0004 & 0.500 $\pm$ < 0.001  \\
       Lacy non-eq. cloudy (thick)  & 274.9 $\pm$ 0.2       & 3.710 $\pm$ 0.025   & 1.068 $\pm$ 0.007 & 2.3598 $\pm$ 0.0005 & -0.288 $\pm$ 0.022 \\
       Lacy non-eq. cloudy (thin)   & 274.3 $\pm$ 0.6       & 3.594 $\pm$  0.004  & 1.058 $\pm$ 0.008 & 1.7734 $\pm$ 0.0005 & -0.168 $\pm$ 0.004  \\
       Sonora Elf Owl               & 306.9 $\pm$ 1.2       & 3.745 $\pm$  0.006  & 0.810 $\pm$ 0.009 & 1.4706 $\pm$ 0.0003  & 0.738 $\pm$ 0.005    \\ 
      \hline    
      \textit{Constrain log(g) to 4.0 - 5.0}\\
      \hline
       ATMO ++ no PH$_3$            & 297.4 $\pm$ 0.6       & 4.166 $\pm$ 0.008   & 0.789 $\pm$ 0.006 & 3.6823 $\pm$ 0.0003  & 0                 \\
       Lacy eq. clear               & 260.7 $\pm$ 0.6       & 4.001 $\pm$ < 0.001 & 0.941 $\pm$ 0.008 & 3.5800 $\pm$ 0.0004  & 0.500 $\pm$ < 0.001  \\
       Lacy eq. cloudy (thick)      & 227.9 $\pm$ 0.2       & 4.000 $\pm$ < 0.001 & 1.539 $\pm$ 0.011 & 9.5625 $\pm$ 0.0010 & -0.180 $\pm$ 0.005 \\ 
       Lacy eq. cloudy (thin)       & 226.9 $\pm$ 0.2       & 4.000 $\pm$ < 0.001 & 1.494 $\pm$ 0.011 & 9.0093 $\pm$ 0.0009  & -0.055 $\pm$ 0.005  \\ 
       Lacy non-eq. clear           & 242.6 $\pm$ 0.4       & 4.248 $\pm$ 0.002 & 1.251 $\pm$ 0.010 & 11.1709 $\pm$ 0.0006 & 0.481 $\pm$ 0.003   \\
       Lacy non-eq. cloudy (thick)  & 229.2 $\pm$ 0.2       & 4.000 $\pm$ < 0.001 & 1.541 $\pm$ 0.011 & 9.5840 $\pm$ 0.0010  & -0.185 $\pm$ 0.002 \\
       Lacy non-eq. cloudy (thin)   & 225.9 $\pm$ 0.1     & 4.000 $\pm$ < 0.001 & 1.576 $\pm$ 0.011 & 10.0277 $\pm$ 0.0010  & -0.072 $\pm$ 0.002 \\
       Sonora Elf Owl               & 275.0 $\pm$ < 0.1     & 4.000 $\pm$ < 0.001 & 0.765 $\pm$ 0.005 & 2.3616 $\pm$ 0.0002  & 0.169 $\pm$ 0.007  \\ 
      \hline    
    \end{tabular}
    \end{table*}

     \begin{table*}
    \caption{The luminosity of the best-fit models calculated by integrating the flux fro 0.45 to 30 µm. We do not calculate the Sonora Elf Owl model as the models only cover up to 15 µm. }             
    \label{table:gridmodres_lum}      
    \centering                          
    \begin{tabular}{l l}        
    \hline\hline                 
    Model &  L/L$_\odot$ [\#] \\    
    \hline                        
       ATMO++ no PH$_3$             & -7.318   \\
       Lacy eq. clear               & -7.458   \\
       Lacy eq. cloudy (thick)      & -7.280   \\
       Lacy eq. cloudy (thin)       & -7.294   \\
       Lacy non-eq. clear           & -7.336   \\ 
       Lacy non-eq. cloudy (thick)  & -7.271   \\
       Lacy non-eq. cloudy (thin)   & -7.271   \\ \hline                                
       Median & -7.294 $\pm$ 0.023 \\ \hline
    \end{tabular}
    \end{table*}
\end{appendix}

\end{document}